\documentclass[prd,nofootinbib,showpacs,preprint]{revtex4}
\usepackage{amsmath}
\usepackage{graphicx}
\usepackage{epsfig}
\graphicspath{{Figs/}}
\usepackage{dcolumn}
\usepackage{bm}
\usepackage{amssymb}
\usepackage[usenames,dvipsnames]{color}
\usepackage{slashed}
\usepackage[dvipdfm,colorlinks,citecolor=blue]{hyperref}
\begin{document}
\title{Radiative Generation of Non-zero $\theta_{13}$ in MSSM \\ with broken $A_4$ Flavor Symmetry}
\author{Manikanta Borah}
\email{mani@tezu.ernet.in}
\author{Debasish Borah}
\email{dborah@tezu.ernet.in}
\author{Mrinal Kumar Das}
\email{mkdas@tezu.ernet.in}
\affiliation{Department of Physics, Tezpur University, Tezpur-784028, India}

\begin{abstract}
We study the renormalization group effects on neutrino masses and mixing in Minimal Supersymmetric Standard Model (MSSM) by considering a $\mu-\tau$ symmetric mass matrix at high energy scale giving rise to Tri-Bi-Maximal (TBM) type mixing. We outline a flavor symmetry model based on $A_4$ symmetry giving rise to the desired neutrino mass matrix at high energy scale. We take the three neutrino mass eigenvalues at high energy scale as input parameters and compute the neutrino parameters at low energy by taking into account of renormalization group effects. We observe that the correct output values of neutrino parameters at low energy are obtained only when the input mass eigenvalues are large $|m_{1,2,3}| = 0.08-0.12$ eV with a very mild hierarchy of either inverted or normal type. A large inverted or normal hierarchical pattern of neutrino masses is disfavored within our framework. We also find a preference towards higher values of $\tan{\beta}$, the ratio of vacuum expectation values (vev) of two Higgs doublets in MSSM in order to arrive at the correct low energy output. Such a model predicting large neutrino mass eigenvalues with very mild hierarchy and large $\tan{\beta}$ could have tantalizing signatures at oscillation, neutrino-less double beta decay as well as collider experiments.

\end{abstract}
\pacs{14.60.Pq, 11.10.Gh, 11.10.Hi}
\maketitle
\section{Introduction}
\label{intro}
Exploration of the origin of neutrino masses and mixing has been one of the major goals of particle physics community for the last few decades.
The results of recent neutrino oscillation experiments have provided a clear evidence favoring the existence of tiny but non-zero 
neutrino masses \cite{PDG}.
Recent neutrino oscillation experiments like T2K \cite{T2K}, Double ChooZ \cite{chooz}, Daya-Bay \cite{daya} and RENO \cite{reno} have not only confirmed 
the earlier predictions for neutrino parameters, but also provided strong evidence for a non-zero value of the reactor mixing angle $\theta_{13}$. 
The latest global 
fit values for 3$\sigma$ range of neutrino oscillation parameters \cite{global} are as follows:

$$ \Delta m_{21}^2=(7.00-8.09) \times 10^{-5} \; \text{eV}^2$$
$$ \Delta m_{31}^2 \;(\text{NH}) =(2.27-2.69)\times 10^{-3} \; \text{eV}^2 $$
$$ \Delta m_{23}^2 \;(\text{IH}) =(2.24-2.65)\times 10^{-3} \; \text{eV}^2 $$
$$ \text{sin}^2\theta_{12}=0.27-0.34 $$
$$ \text{sin}^2\theta_{23}=0.34-0.67 $$ 
\begin{equation}
\text{sin}^2\theta_{13}=0.016-0.030
\label{equation:data}
\end{equation}
where NH and IH refers to normal and inverted hierarchy respectively. Another global fit study \cite{global0} reports the 3$\sigma$ values as
$$ \Delta m_{21}^2=(6.99-8.18) \times 10^{-5} \; \text{eV}^2$$
$$ \Delta m_{31}^2 \;(\text{NH}) =(2.19-2.62)\times 10^{-3} \; \text{eV}^2 $$
$$ \Delta m_{23}^2 \;(\text{IH}) =(2.17-2.61)\times 10^{-3} \; \text{eV}^2 $$
$$ \text{sin}^2\theta_{12}=0.259-0.359 $$
$$ \text{sin}^2\theta_{23}=0.331-0.637 $$ 
\begin{equation}
\text{sin}^2\theta_{13}=0.017-0.031
\label{equation:data2}
\end{equation}

The observation of non-zero $\theta_{13}$ which is evident from the above global fit data can have non-
trivial impact on 
neutrino mass hierarchy as studied in recent papers \cite{mkd}. Non-zero $\theta_{13}$ can also shed light on the Dirac CP violating phase in the leptonic sector which would have remained unknown if $\theta_{13}$ were exactly zero. The detailed analysis of this non-zero $\theta_{13}$  have been demonstrated both from 
theoretical \cite{zee1}, as well as phenomenological \cite{king} point of view, prior to and after the confirmation of this important result announced in
2012. It should be noted that prior to the discovery of non-zero $\theta_{13}$, the neutrino oscillation data were compatible with the so called TBM form of the neutrino mixing matrix \cite{Harrison} given by
\begin{equation}
U_{TBM}==\left(\begin{array}{ccc}\sqrt{\frac{2}{3}}&\frac{1}{\sqrt{3}}&0\\
 -\frac{1}{\sqrt{6}}&\frac{1}{\sqrt{3}}&\frac{1}{\sqrt{2}}\\
\frac{1}{\sqrt{6}}&-\frac{1}{\sqrt{3}}& \frac{1}{\sqrt{2}}\end{array}\right),
\end{equation}
which predicts $\text{sin}^2\theta_{12}=\frac{1}{3}$, $\text{sin}^2\theta_{23}=\frac{1}{2}$ and $\text{sin}^2\theta_{13}=0$. However, since the latest data have ruled out $\text{sin}^2\theta_{13}=0$, there arises the need to go beyond the TBM framework. In view of the importance of the non-zero reactor mixing and hence, CP violation in neutrino sector, the present work demonstrates how a specific $\mu$-$\tau$ symmetric mass matrix (giving rise to TBM type mixing) at high energy scale can produce non-zero $\theta_{13}$ along with the desired values of other neutrino parameters $\Delta m_{21}^2, \Delta m_{23}^2, \theta_{23}, \theta_{12} $ at low energy scale through renormalization group evolution (RGE). We also outline how the $\mu-\tau$ symmetric neutrino mass matrix with TBM type mixing can be realized at high energy scale within the framework of MSSM with an additional $A_4$ flavor symmetry at high energy scale. After taking the RGE effects into account, we observe that the output at TeV scale is very much sensitive to the choice of neutrino mass ordering at high scale as well as the value of $\text{tan}{\beta} = \frac{v_u}{v_d}$, the ratio of vev's of two MSSM Higgs doublets $H_{u,d}$. We point out that this model allows only a very mild hierarchy of both inverted and normal type at high energy scale. We scan the neutrino mass eigenvalues at high energy and constrain them to be large $|m_{1,2,3}| = 0.08-0.12$ eV in order to produce correct neutrino parameters at low energy. We consider two such input values for mass eigenvalues, one with inverted hierarchy and the other with normal hierarchy and show the predictions for neutrino parameters at low energy scale. We also show the evolution of effective neutrino mass $m_{ee} = \lvert \sum_i U^2_{ei} m_i \rvert$ (where $U$ is the neutrino mixing matrix) that could be interesting from neutrino-less double beta decay point of view. Finally we consider the cosmological upper bound on the sum of absolute neutrino masses ($\sum_i \lvert m_i \rvert < 0.23$ eV) reported by the Planck collaboration \cite{planck} to check if the output at low energy satisfy this or not.

This article is organized as follows. In section \ref{model}, we discuss briefly the $A_4$ model at high energy scale. In section \ref{rge} we outline the RGE's 
of mass eigenvalues and mixing parameters. In section \ref{result} we discuss our numerical results, and finally conclude in  section \ref{conclude}.

\section{$A_4$ model for neutrino mass}
\label{model}
Type I seesaw framework is the  simplest mechanism for generating tiny neutrino masses and mixing. In this seesaw mechanism neutrino mass matrix 
can be written as
\begin{equation}
m_{LL}=-m_{LR}M_R^{-1}m_{LR}^T.
\end{equation}
Within this framework of seesaw mechanism neutrino mass has been extensively
studied by discrete flavor groups by many authors \cite{ish} available in the literature. Among the different discrete groups the model by the 
finite group of even permutation, $A_4$   
also can explain
the $\mu-\tau$ symmetric mass matrix obtained from this type I seesaw mechanism. This group has $12$ elements having $4$ irreducible representations, with 
dimensions $n_i$, such that $\sum_i n_i^2=12$. The characters of $4$ representations are shown in table \ref{table:character}. The complex number $\omega$ 
is the cube root of unity. In the present work we outline a neutrino mass model with $A_4$ symmetry given in the ref.\cite{alt1}. This flavor symmetry is also accompanied by an additional $Z_3$ symmetry in order to achieve the desired leptonic mixing. 
In this model, the three families of left-handed lepton 
doublets $l = (l_e, l_{\mu}, l_{\tau})$ transform as triplets,
while the electroweak singlets $e^c$, $\mu^c$, $\tau^c$ and the electroweak Higgs doublets $H_{u,d}$ transform as
singlets under the $A_4$ symmetry. In order to break the flavor symmetry spontaneously, two $A_4$ triplet scalars $\phi_l=(\phi_{l1}, \phi_{l2}, \phi_{l3}), \phi_{\nu}=(\phi_{\nu1}, \phi_{\nu2}, \phi_{nu3})$ and three scalars $\zeta_1, \zeta_2, \zeta_3$ transforming as $\bf{1}, \bf{1'}, \bf{1''}$ under $A_4$ are introduced. The $Z_3$ charges for $l, H_{u,d}, \phi_l, \phi_{\nu}, \zeta_{1,2,3}$ are $\omega, 1, 1, \omega, \omega$ respectively. 
\begin{table}[ht]
\centering
\caption{Character table of $A_4$}
\vspace{0.5cm}
\begin{tabular}{ccccc}
 \hline
   Class & $\chi^{(1)}$ &  $\chi^{(2)}$ &  $\chi^{(3)}$&  $\chi^{(4)}$\\ \hline \hline
   $C_1$ & 1&1&1&3 \\ \hline 
    $C_2$& 1&$\omega$&$\omega^2$&0 \\  
$C_3$ &1&$\omega^2$&$\omega$&0\\ 
$C_4$ &1&1&1&-1\\  \hline
\end{tabular}
\label{table:character}
\end{table}

Under the electroweak gauge symmetry as well as the flavor symmetry mentioned above, the superpotential for the neutrino sector can be written as
\begin{equation}
W_{\nu} = (y_{\nu \phi} \phi_{\nu} +y_{\nu \zeta1} \zeta_1 + y_{\nu \zeta2} \zeta_2 + y_{\nu \zeta3} \zeta_3 )\frac{llH_uH_u}{\Lambda^2}
\end{equation}
where $\Lambda$ is the cutoff scale and $y'$s are dimensionless couplings. Decomposing the first term (which is in a $\bf{3}\times \bf{3} \times \bf{3}$ form of $A_4$) into $A_4$ singlets, we get
$$ ll\phi_{\nu} =(2l_el_e-l_{\mu}l_{\tau}-l_{\tau}l_{\mu})\phi_{\nu1}+(2l_{\mu}l_{\mu}-l_{e}l_{\tau}-l_{\tau}l_{e})\phi_{\nu2}+(2l_{\tau}l_{\tau}-l_{e}l_{\mu}-l_{\mu}l_{e})\phi_{\nu3}$$
Similarly, the decomposition of the last three terms into $A_4$ singlet gives
$$ ll\zeta_1 = (l_el_e+l_{\mu}l_{\tau}+l_{\tau}l_{\mu}) \zeta_1 $$
$$ ll\zeta_2 = (l_{\mu}l_{\mu}+l_{e}l_{\tau}+l_{\tau}l_{e}) \zeta_2 $$
$$ ll\zeta_3 = (l_{\tau}l_{\tau}+l_{e}l_{\mu}+l_{\mu}l_{e}) \zeta_3 $$
Assuming the vacuum alignments of the scalars as $\langle \phi_{\nu} \rangle = \alpha_{\nu}\Lambda (1,1,1), \langle \zeta_1 \rangle = \alpha_{\zeta} \Lambda, \langle \zeta_{2,3} \rangle = 0$, the neutrino mass matrix can be written as
\begin{equation}
\label{matrix1}
 m_{LL}=\frac{v_u^2}{\Lambda}\left(\begin{array}{ccc}
 a+2d/3&-d/3&-d/3\\
 -d/3&2d/3&a-d/3\\
 -d/3&a-d/3&2d/3
 \end{array}\right),
\end{equation}
where $d = y_{\nu \phi} \alpha_{\nu}, a = y_{\nu \zeta1} \alpha_{\zeta}$ and $v_u$ is the vev of $H_u$. The above mass matrix has eigenvalues $m_1=\frac{v_u^2}{\Lambda}(a+d)$, $m_2=\frac{v_u^2}{\Lambda}a$ and $m_3=\frac{v_u^2}{\Lambda}(-a+d)$. Without adopting any un-natural fine tuning condition to relate the mass eigenvalues further, we wish to keep all the three neutrino mass eigenvalues as free parameters in the $A_4$ symmetric theory at high energy and determine the most general parameter space at high energy scale which can reproduce the correct neutrino oscillation data at low energy through renormalization group evolution (RGE).

Such a parameterization of the neutrino mass matrix however, does not disturb the generic features of 
the model for example, the $\mu-\tau$ symmetric nature of $m_{LL}$, TBM type mixing as well the diagonal nature of the charged lepton mass matrix, which at leading order (LO) is given by \cite{alt1,galt}
\begin{equation}
 m_{l}=v_d \alpha_l\left(\begin{array}{ccc}
 y_e&0&0\\
 0&y_{\mu}&0\\
 0&0&y_{\tau}
 \end{array}\right)
 \label{clepton}
\end{equation}

Here $v_d$ is the vev of $H_d$; $y_e$, $y_\mu$, $y_\tau$ and $\alpha_l$ are dimensionless couplings. These matrices in the leptonic sector given by (\ref{matrix1}) and (\ref{clepton}) are used in the next section for numerical analysis.
\begin{center} 
\begin{figure}
\includegraphics[width=1.0\textwidth]{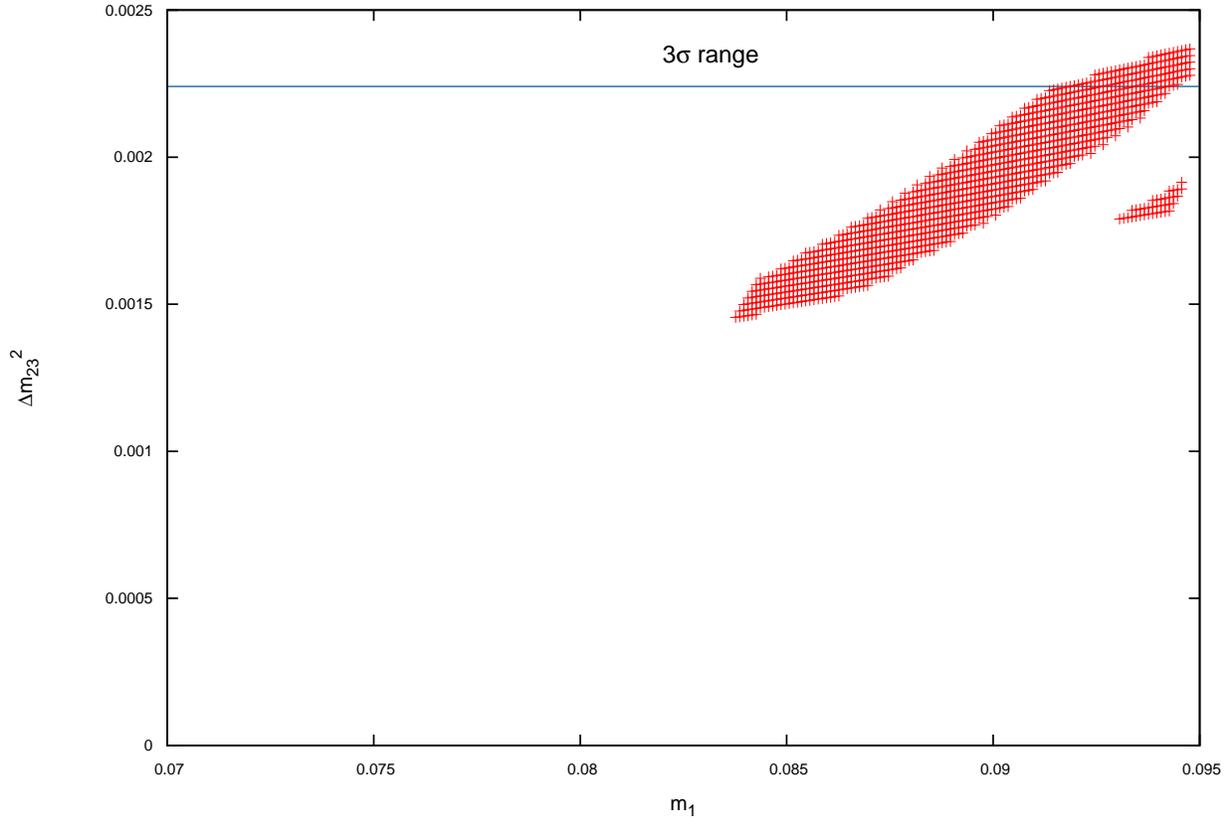}
\caption{Scatter plot showing $\Delta m_{23}^2$ at low energy versus initial value of $m_1$ at high energy for inverted hierarchy while keeping all other neutrino parameters at low energy within $3\sigma$ range}
\label{fig15}
\end{figure}
\end{center}
\begin{center} 
\begin{figure}
\includegraphics[width=1.0\textwidth]{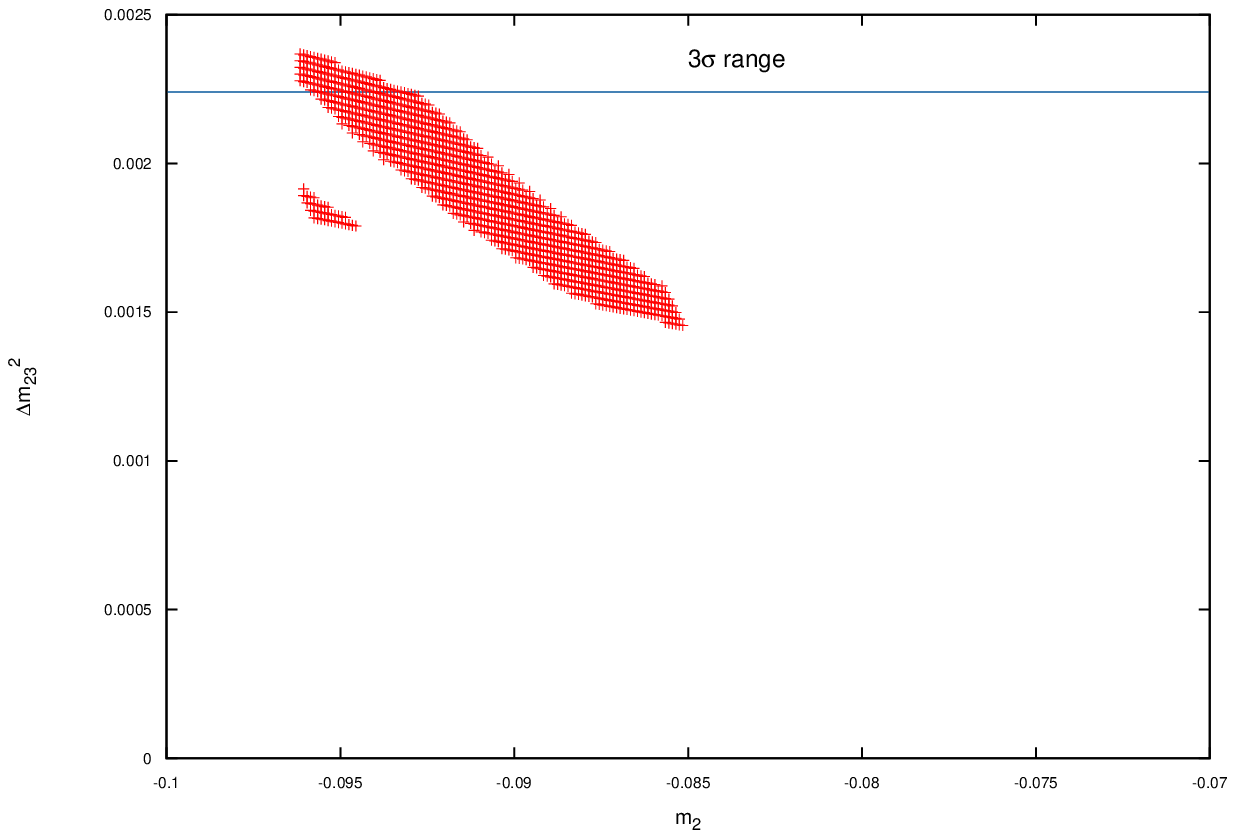}
\caption{Scatter plot showing $\Delta m_{23}^2$ at low energy versus initial value of $m_2$ at high energy for inverted hierarchy while keeping all other neutrino parameters at low energy within $3\sigma$ range}
\label{fig16}
\end{figure}
\end{center}
\begin{center} 
\begin{figure}
\includegraphics[width=1.0\textwidth]{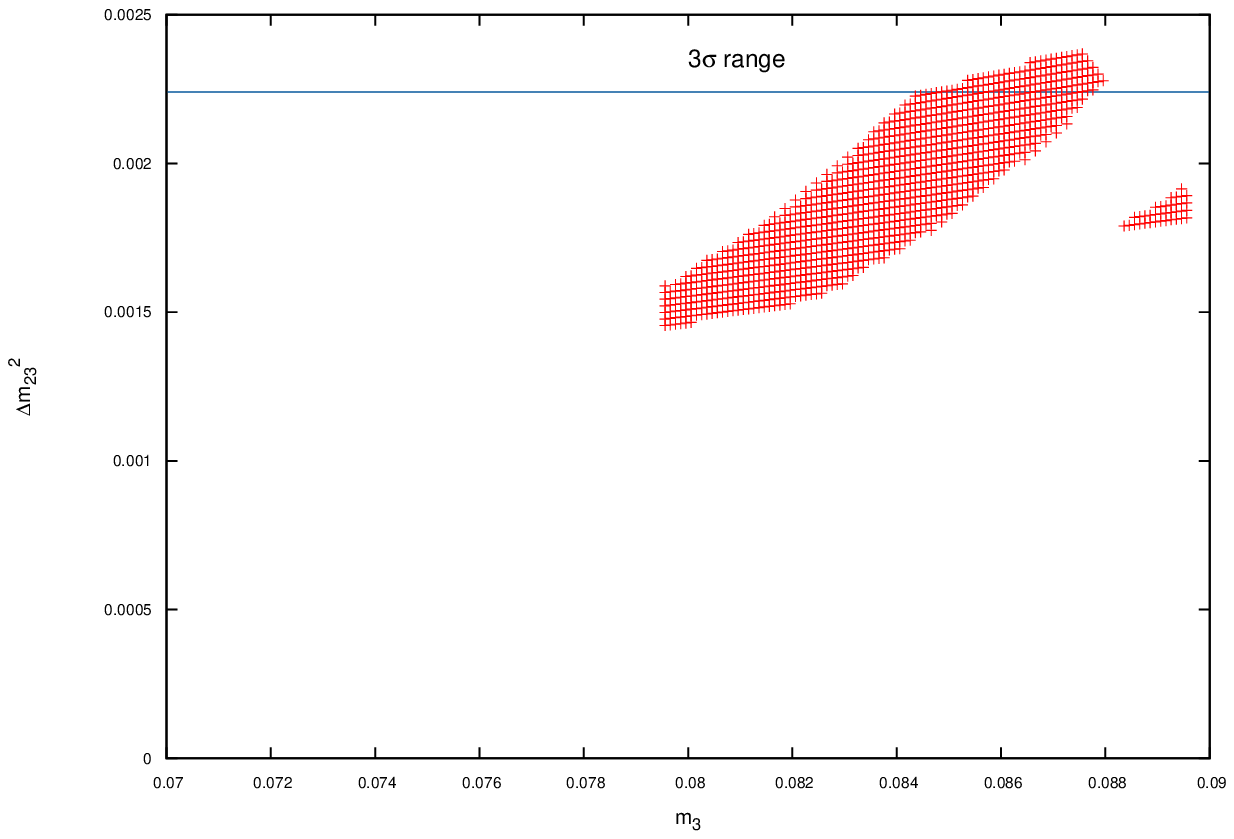}
\caption{Scatter plot showing $\Delta m_{23}^2$ at low energy versus initial value of $m_3$ at high energy for inverted hierarchy while keeping all other neutrino parameters at low energy within $3\sigma$ range}
\label{fig17}
\end{figure}
\end{center}
\begin{center} 
\begin{figure}
\includegraphics[width=1.0\textwidth]{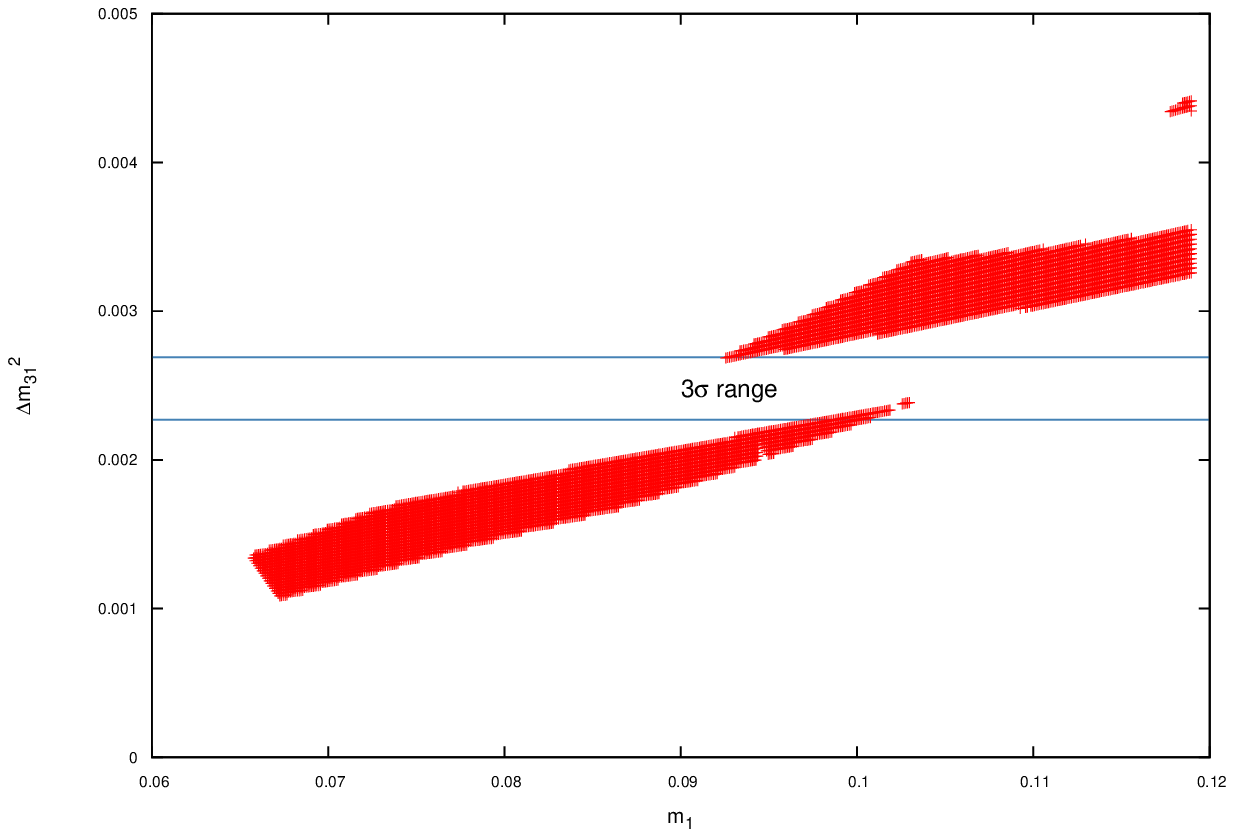}
\caption{Scatter plot showing $\Delta m_{31}^2$ at low energy versus initial value of $m_1$ at high energy for normal hierarchy while keeping all other neutrino parameters at low energy within $3\sigma$ range}
\label{fig18}
\end{figure}
\end{center}
\begin{center} 
\begin{figure}
\includegraphics[width=1.0\textwidth]{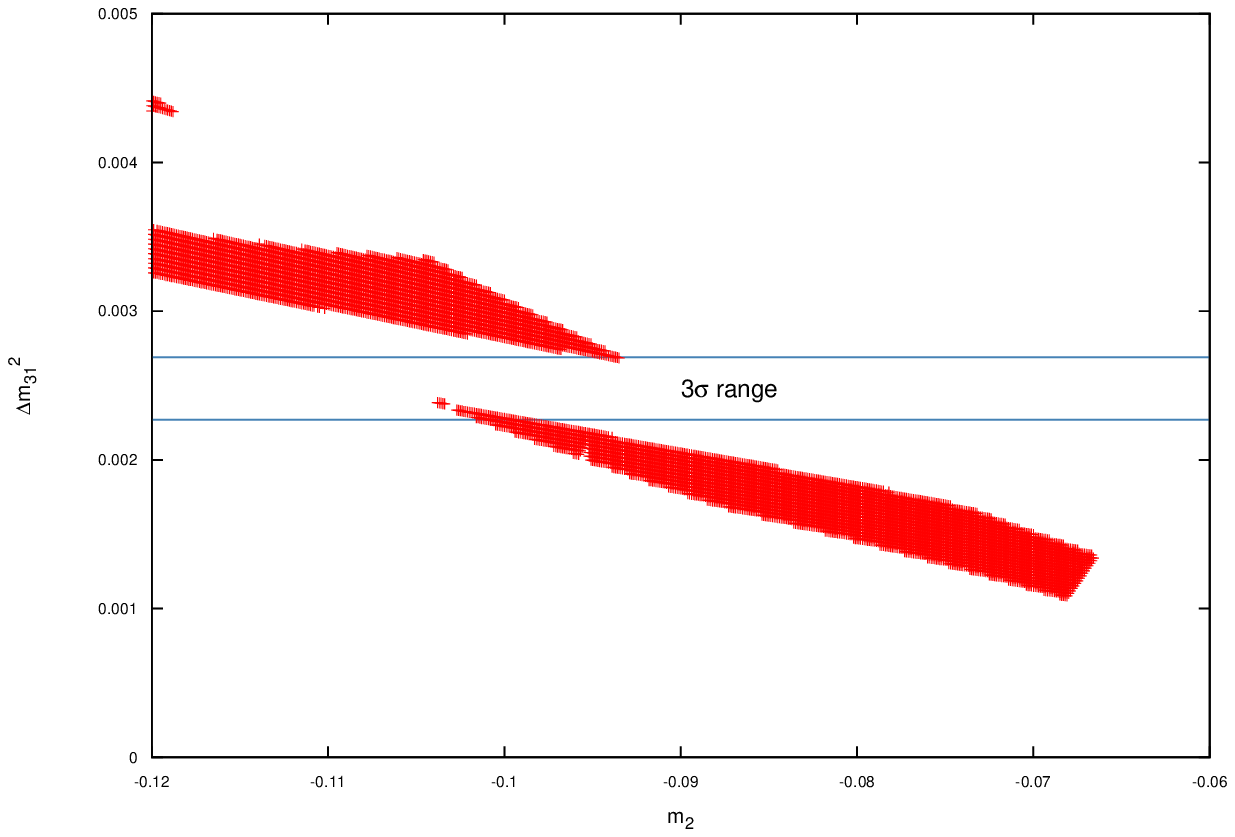}
\caption{Scatter plot showing $\Delta m_{31}^2$ at low energy versus initial value of $m_2$ at high energy for normal hierarchy while keeping all other neutrino parameters at low energy within $3\sigma$ range}
\label{fig19}
\end{figure}
\end{center}
\begin{center} 
\begin{figure}
\includegraphics[width=1.0\textwidth]{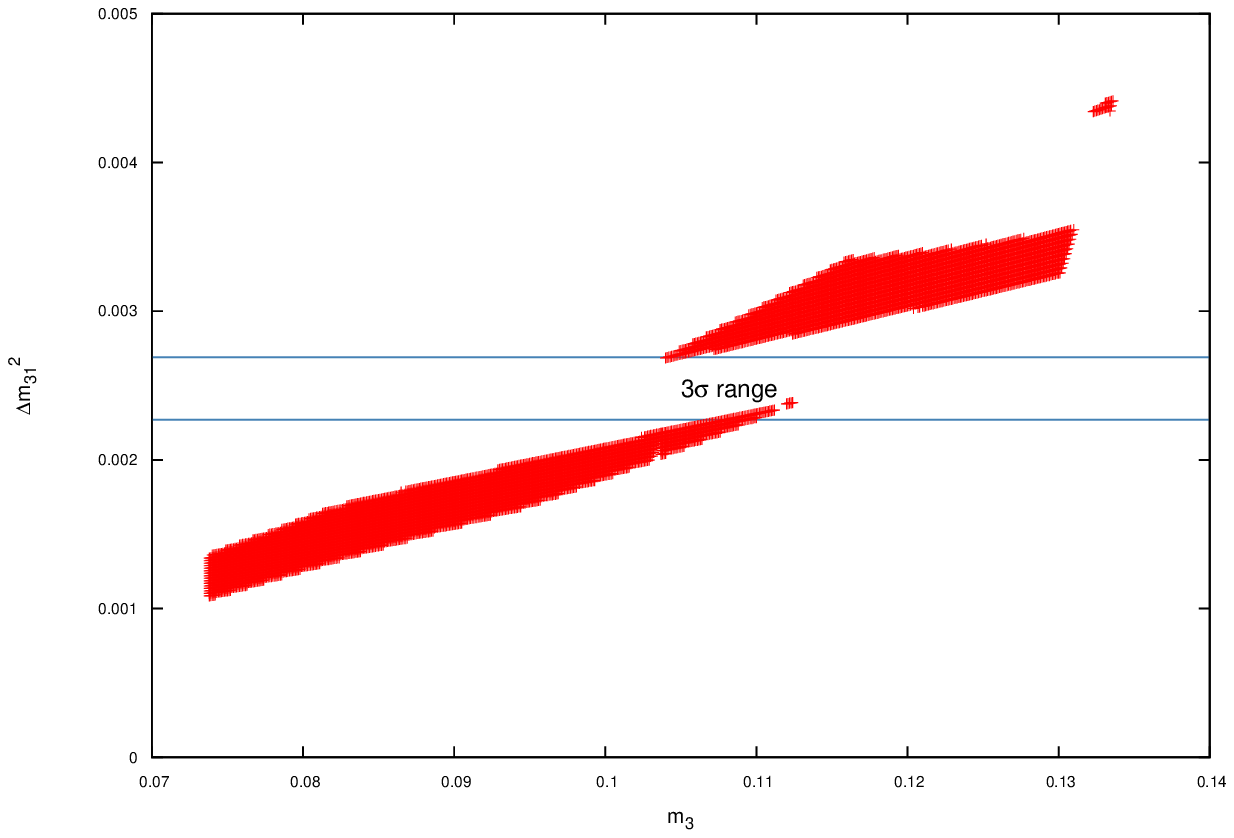}
\caption{Scatter plot showing $\Delta m_{31}^2$ at low energy versus initial value of $m_3$ at high energy for normal hierarchy while keeping all other neutrino parameters at low energy within $3\sigma$ range}
\label{fig20}
\end{figure}
\end{center}
\begin{center} 
\begin{figure}
\includegraphics[width=1.0\textwidth]{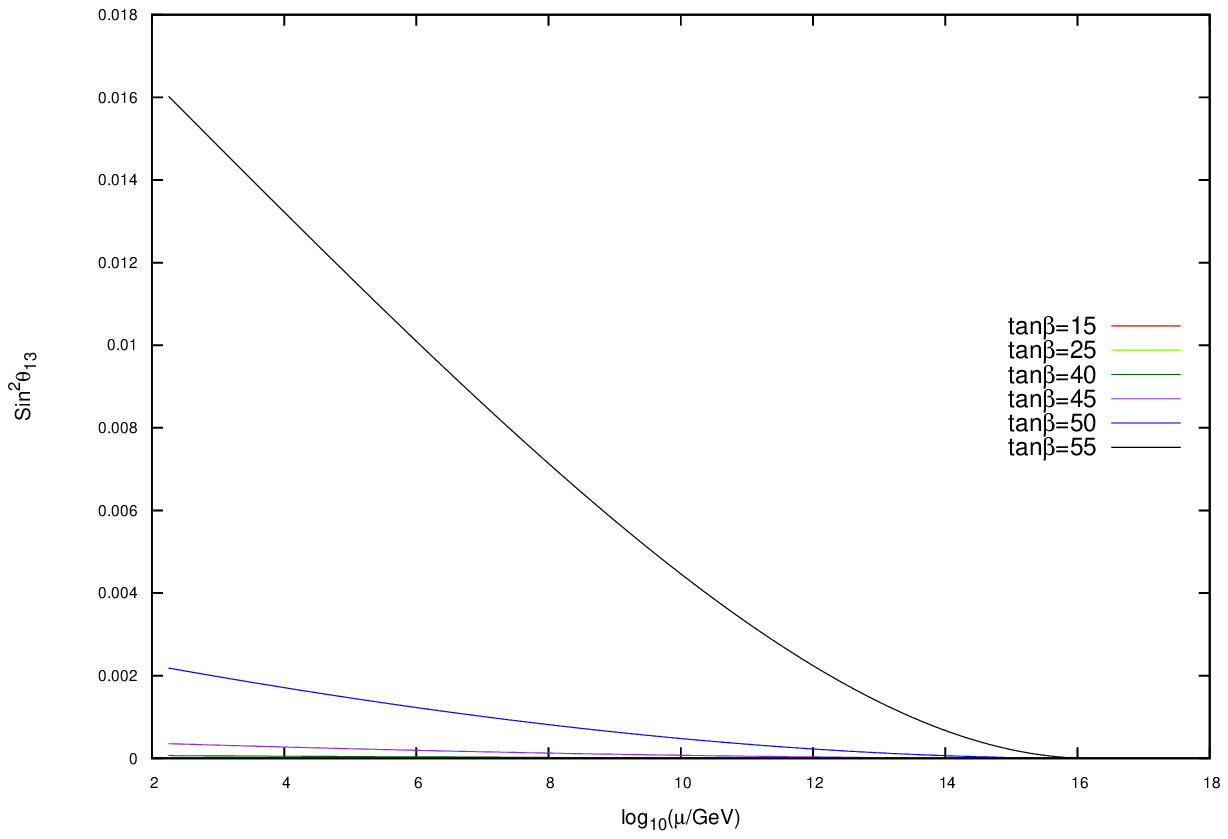}
\caption{Radiative generation of $\text{sin}^2\theta_{13}$ for
 tan$\beta$=15, 25, 40, 45, 50, 55 for inverted hierarchy using input values given in Table \ref{values1}}
\label{fig1}
\end{figure}
\end{center}
\begin{center} 
\begin{figure}
\includegraphics[width=1.0\textwidth]{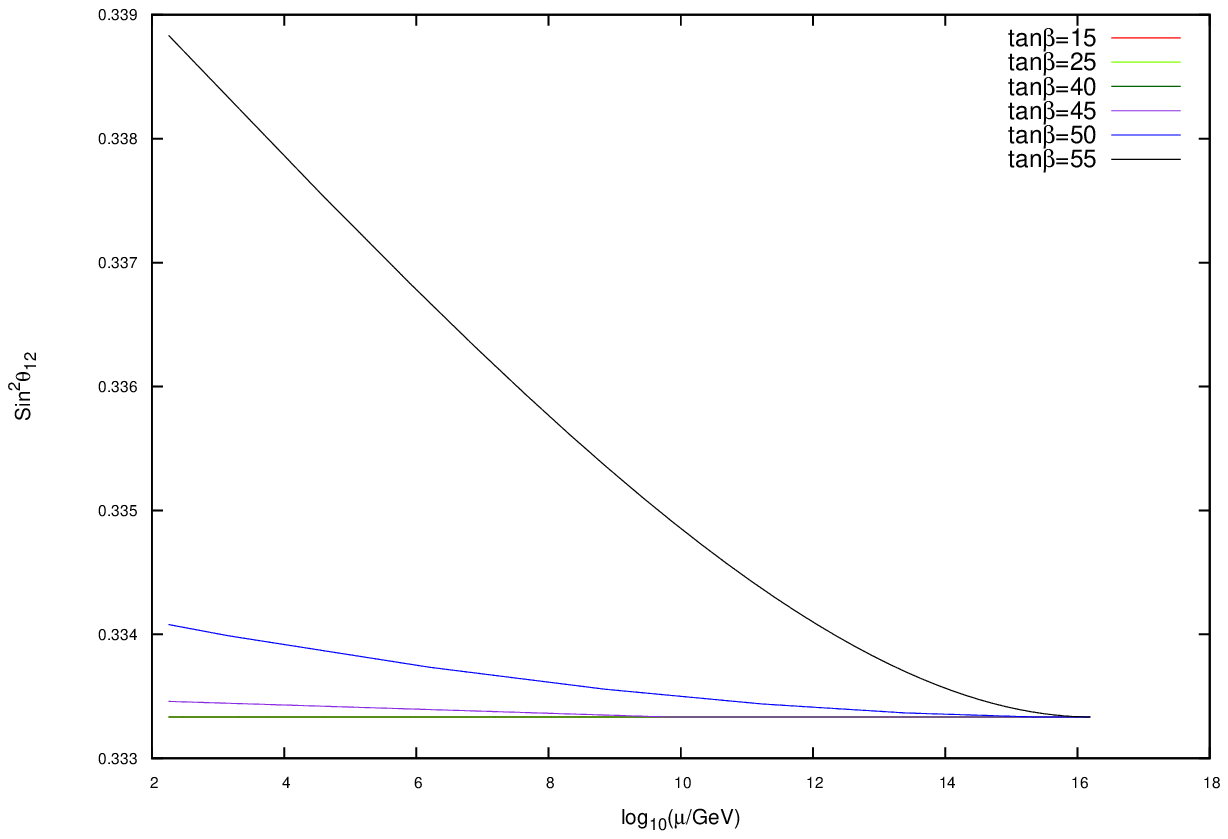}
\caption{Evolution of $\text{sin}^2\theta_{12}$ for
 tan$\beta$=15, 25, 40, 45, 50, 55 for inverted hierarchy using input values given in Table \ref{values1}}
\label{fig2}
\end{figure}
\end{center}
\section{RGE for neutrino masses and mixing}
\label{rge}
The left-handed Majorana neutrino mass matrix $m_{LL}$ which is generally obtained from seesaw mechanism at high scale $M_R$, 
is usually  expressed in terms of $K(t)$, the coefficient of the dimension five neutrino mass operator \cite{pc, sa} in a scale-dependent manner \cite{nns},
\begin{equation}
m_{LL}(t)=v^2_{u}K(t),
\label{evom}
\end{equation}
where $t=\ln(\mu/1GeV)$ and the vev is $v_{u}=v_0 \sin\beta$ with $v_0=174$ GeV in MSSM. The neutrino mass eigenvalues $m_i$  and the Pontecorvo-Maki-Nakagawa-Sakata (PMNS) mixing matrix $U_{PMNS}$ \cite{bp} are then extracted  through the diagonalization of $m_{LL}(t)$ at every
point in the energy scale $t$ using the equations (\ref{evom}),
\begin{equation}
m_{LL}^{\text{diag}}=\text{diag}(m_{1},m_{2},m_{3})=V^T_{\nu L}m_{LL}V_{\nu L},
\end{equation}
and $U_{PMNS}=V_{\nu L}$ in the basis where the charged lepton mass 
matrix is diagonal. The PMNS mixing matrix,   
\begin{equation}
U_{PMNS}=\left(\begin{array}{ccc}
U_{e1} & U_{e2} & U_{e3} \\
U_{\mu 1} & U_{\mu 2} & U_{\mu 3} \\
U_{\tau 1} & U_{\tau 2} & U_{\tau 3}
\end{array}\right),
\end{equation}
 is usually  parameterized in terms  of the product of three rotations  $R(\theta_{23})$, $R(\theta_{13})$ and  $R(\theta_{12})$,
 (neglecting CP violating phases) by 
\begin{equation}
U_{PMNS}=U_l^{\dagger}U_{\nu}=\left(\begin{array}{ccc}
c_{13}c_{12}  & c_{13}s_{12}  & s_{13} \\
-c_{23}s_{12}-c_{12}s_{13}s_{23}  & c_{12}c_{23}-s_{12}s_{13}s_{23} & c_{13}s_{23} \\
s_{12}s_{23}-c_{12}s_{13}c_{23} & -c_{12}s_{23}-c_{23}s_{13}s_{12} & c_{13}c_{23}
\end{array}\right),
\end{equation}
where $U_l$ is unity in the basis where charge lepton mass matrix is diagonal,  $s_{ij}=\sin{\theta_{ij}}$ and $c_{ij}=\cos{\theta_{ij}}$  respectively.

The RGE's for $v_u$ and the eigenvalues of coefficient $K(t)$ in equation (\ref{evom}), defined in the basis
 where the charged lepton mass matrix is diagonal, can be expressed as \cite{ph,nns1}
\begin{equation}
\frac{d}{dt}\text{ln}v_u = \frac{1}{16\pi^2}[\frac{3}{20}g_1^2+\frac{3}{4}g_2^2-3h_t^2]
\end{equation}
\begin{equation}
\frac{d}{dt}\text{ln}K=-\frac{1}{16\pi^2}[\frac{6}{5}g_{1}^{2}+6g_{2}^2-6h_t^2-\delta_{i3}h_{\tau}^2-\delta_{3j}h_{\tau}^2]
\end{equation}
Neglecting $h_{\mu}^2$ and $h_{e}^2$ compared to $h^2_{\tau}$, and taking scale-independent vev as in equation (\ref{evom}), we have the complete RGE's
 for three neutrino mass eigenvalues, 
\begin{equation}
\frac{d}{dt}m_{i}=\frac{1}{16\pi^2}[(-\frac{6}{5}g_{1}^{2}-6g_{2}^2+6h_{t}^2)+2h_{\tau}^{2}U_{\tau i}^{2}]m_{i}.
\label{masseigen}
\end{equation}
The above equations together with the evolution equations for mixing angles (\ref{angle1}-\ref{angle2}), are  used for the numerical analysis in our work. 

The  approximate analytical solution of equation (\ref{masseigen}) can be obtained  by taking static mixing angle $U_{\tau i}^2$ in the integration range
as \cite{mkp}
\begin{equation}
m_{i}(t_{0})=m_{i}(t_R)exp(\frac {6}{5} I_{g1}+6I_{g2}-6I_{t})exp(-2U_{\tau i}^2 I_{\tau})
\label{mass}
\end{equation} 
The integrals in the above expression are usually defined as\cite{nns, mkp} 
\begin{equation}
I_{gi}(t_{0})=\frac{1}{16\pi^2}\int_{t_{0}}^{t_{R}}g_{i}^{2}(t)dt
\end{equation}
and
\begin{equation}
I_{f}(t_{0})=\frac{1}{16\pi^2}\int_{t_{0}}^{t_{R}}h_{f}^{2}(t)dt
\end{equation}
where $i=1,2,3$ and $f=t,b,\tau$ respectively.
For a two-fold degenerate neutrino masses that is,
 $m_{LL}^{diag}= \text{diag}(m, m, m')=U_{PMNS}^T m_{LL}U_{PMNS}$, the equation (\ref{mass}) is further  simplified to the following expressions
\begin{equation}
m_{1}(t_0)\approx m(t_R)(1+2\delta_{\tau}(c_{12}s_{13}c_{23}-s_{12}s_{23})^2)+O(\delta_{\tau}^2)
\label{m1}
\end{equation}
\begin{equation}
m_{2}(t_0)\approx m(t_R)(1+2\delta_{\tau}(c_{23}s_{13}s_{12}+c_{12}s_{23})^2)+O(\delta_{\tau}^2)
\label{m2}
\end{equation}
\begin{equation}
m_{3}(t_0)\approx m'(t_R)(1+2\delta_{\tau}(c_{13}c_{23})^2)+O(\delta_{\tau}^2).
\label{m3}
\end{equation} 
While deriving the above expressions, the following approximations are used
$$ \exp(-2|U_{\tau i}|^2 I_{\tau})\simeq 1-2|U_{\tau i}|^2 I_{\tau}=1+2|U_{\tau i}|^2 \delta_{\tau}$$
$$ -\delta_{\tau}=I_{\tau} \simeq \frac{1}{\cos ^2\beta}(m_{\tau}/4 \pi v)^2 \ln (M_R/m_t)$$

The sign of the quantity  $\delta_{\tau}$ in MSSM depends on the neutrino mixing matrix parameters and the approximation on $\delta_{\tau}$
taken here is valid only if $t_{0}$ is associated with the top quark mass. From equations (\ref{m1}) and (\ref{m2}), 
 the low energy solar neutrino mass scale is then obtained as 
\begin{equation}
\bigtriangleup m^2_{21}(t_0)= m^2_{2}-m^2_{1}\approx 
4 \delta_{\tau}m^2(\cos 2\theta_{12}(s^2_{23}-s^2_{13}c^2_{23})+s_{13}\sin 2\theta_{12}\sin 2\theta_{23})+O(\delta^2_{\tau})
\label{solar}
\end{equation}
\begin{center} 
\begin{figure}
\includegraphics[width=1.0\textwidth]{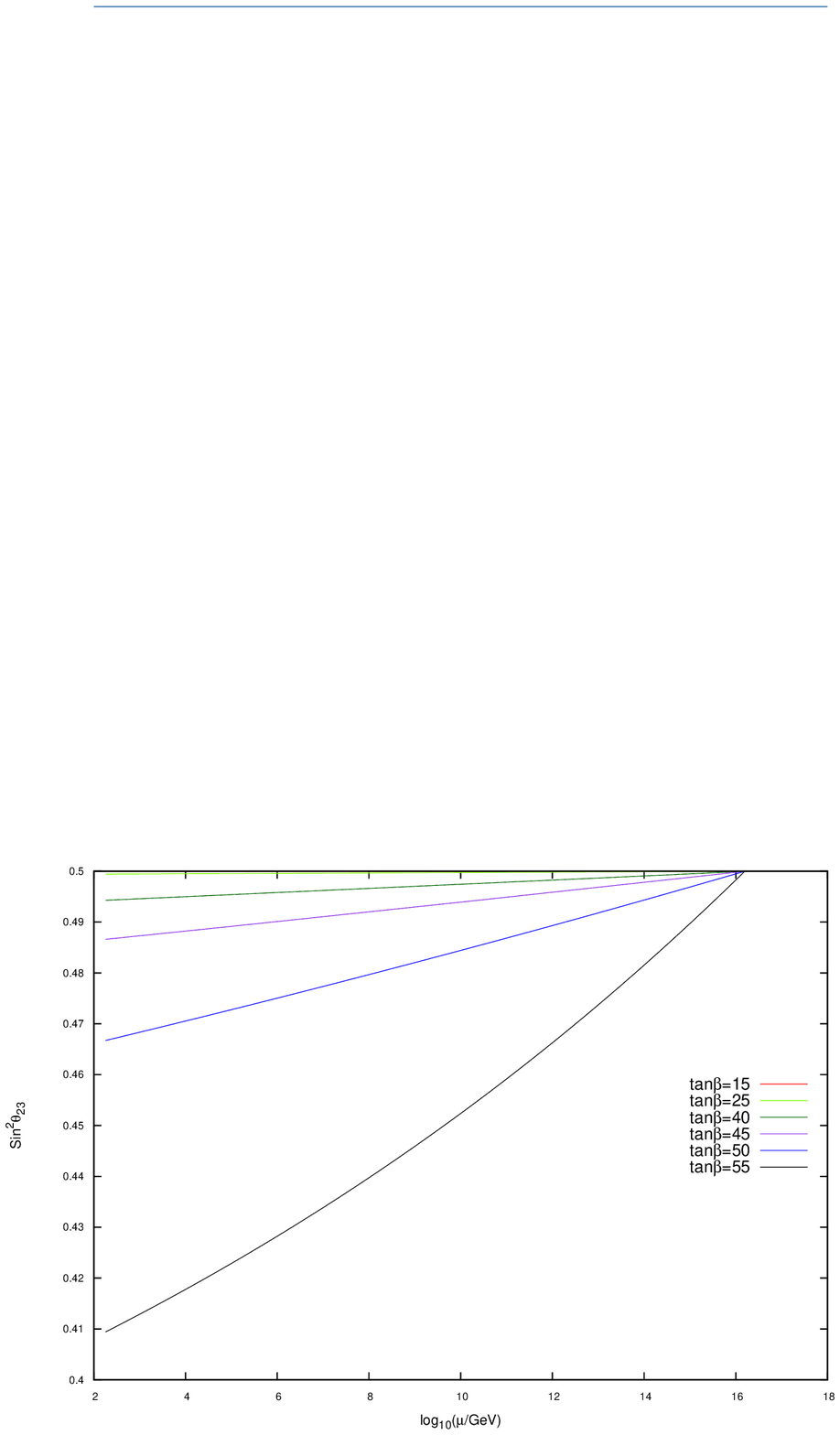}
\caption{Evolution of $\text{sin}^2\theta_{23}$ for
 tan$\beta$=15, 25, 40, 45, 50, 55 for inverted hierarchy using input values given in Table \ref{values1}}
\label{fig3}
\end{figure}
\end{center}
\begin{center} 
\begin{figure}
\includegraphics[width=1.0\textwidth]{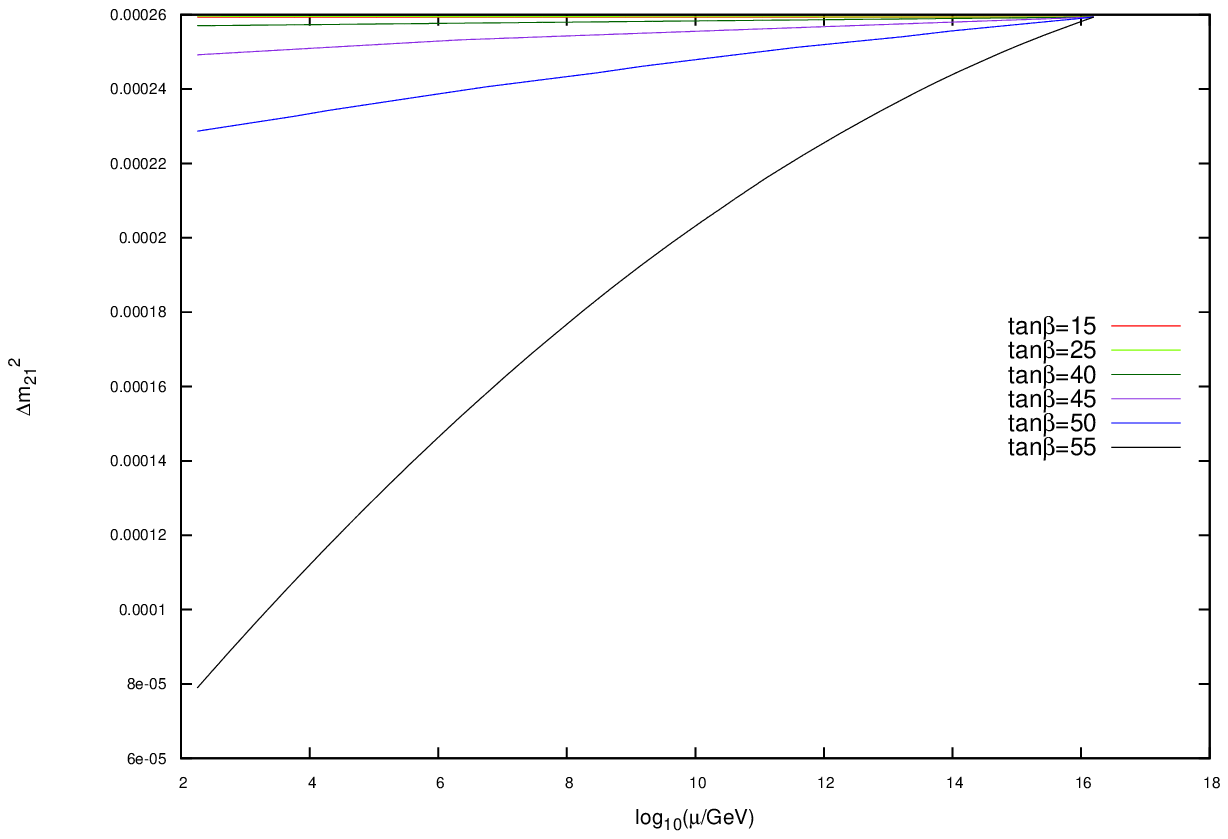}
\caption{Evolution of $\Delta m_{21}^2$ for
 tan$\beta$=15, 25, 40, 45, 50, 55 for inverted hierarchy using input values given in Table \ref{values1}}
\label{fig4}
\end{figure}
\end{center}
 
\subsection{Evolution equations for mixing angles}

The corresponding evolution equations for the PMNS matrix elements $U_{fi}$ are given by \cite{ph}
\begin{equation}
\frac{d U_{fi}}{dt}=-\frac{1}{16\pi^2}\sum_{k\neq i}\frac{m_{k}+m_{i}}{m_{k}-m_{i}} U_{fk}(U^TH^2_eU)_{ki},
\label{um}
\end{equation}
where $f=e,\mu,\tau$ and $i,k=1,2,3$ respectively. Here $H_e$ is the Yukawa coupling 
matrices of the charged leptons in the diagonal basis and 
$$(U^{T}H_e^2U)_{ki}=h_{\tau}^2(U_{k\tau}^{T}U_{\tau i})+
h_{\mu}^2(U_{k\mu}^{T}U_{\mu i})+h_{e}^2(U_{ke}^{T}U_{e i})$$
Neglecting $h_{\mu}^2$ and $h_{e}^2$ as before and denoting $A_{ki}=\frac{m_{k}+m_{i}}{m_{k}-m_{i}}$, 
equation (\ref{um}) simplifies to \cite{ph}
\begin{equation}
\frac{d s_{12}}{dt}=\frac{1}{16\pi^2}h_{\tau}^2 c_{12}[c_{23}s_{13}s_{12}U_{\tau 1}A_{31}-c_{23}s_{13}c_{12}U_{\tau 2}A_{32}
+U_{\tau 1}U_{\tau 2} A_{21}],
\label{angle1}
\end{equation}

\begin{equation}
\frac{d s_{13}}{dt}=\frac{1}{16\pi^2}h_{\tau}^2 c_{23}c_{13}^2[c_{12}U_{\tau 1}A_{31}+s_{12}U_{\tau 2}A_{32}],
\label{angle2}
\end{equation}

\begin{equation}
\frac{d s_{23}}{dt}=\frac{1}{16\pi^2}h_{\tau}^2 c_{23}^2[-s_{12}U_{\tau 1}A_{31}+c_{12}U_{\tau 2}A_{32}].
\label{angle3}
\end{equation}
These equations are valid for a generic MSSM with the minimal field content and are independent of the flavor symmetry structure at high energy scale.

\begin{center} 
\begin{figure}
\includegraphics[width=1.0\textwidth]{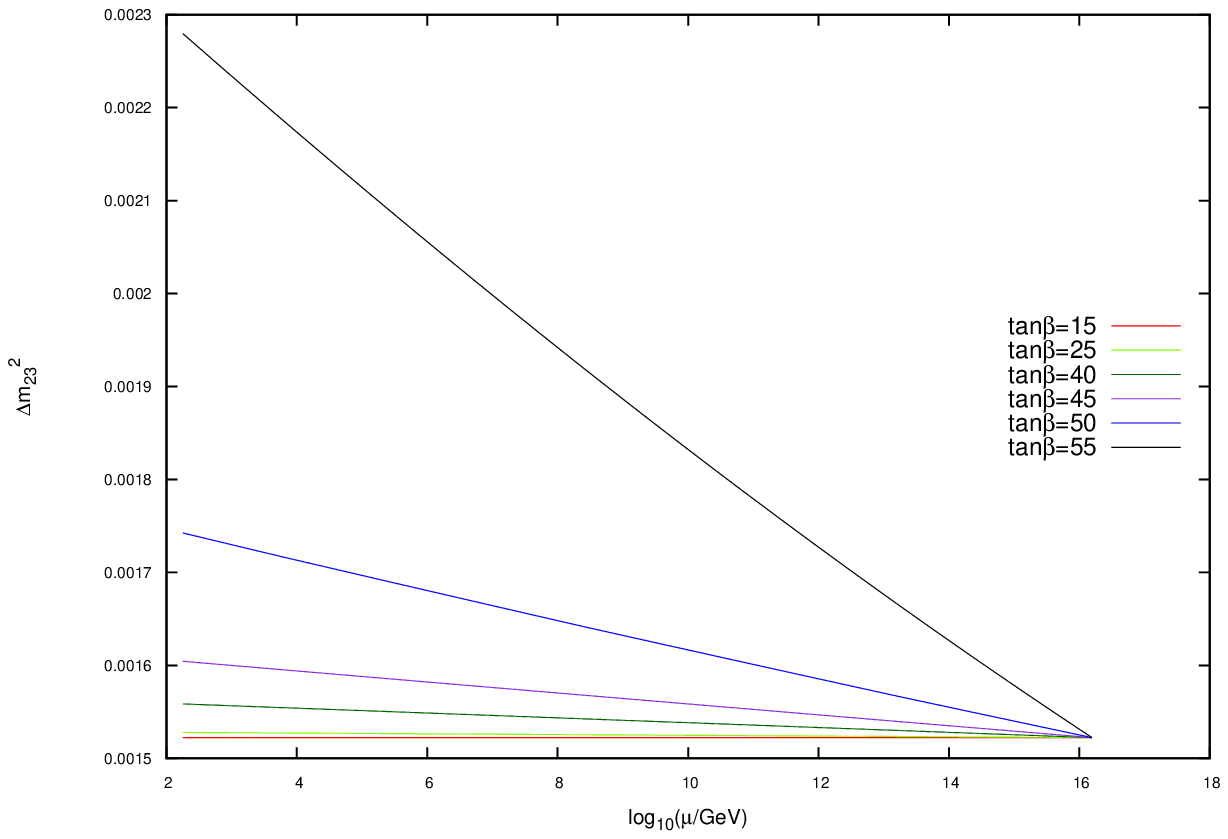}
\caption{Evolution of $\Delta m_{23}^2$ for
 tan$\beta$=15, 25, 40, 45, 50, 55 for inverted hierarchy using input values given in Table \ref{values1}}
\label{fig5}
\end{figure}
\end{center}
\begin{center} 
\begin{figure}
\includegraphics[width=1.0\textwidth]{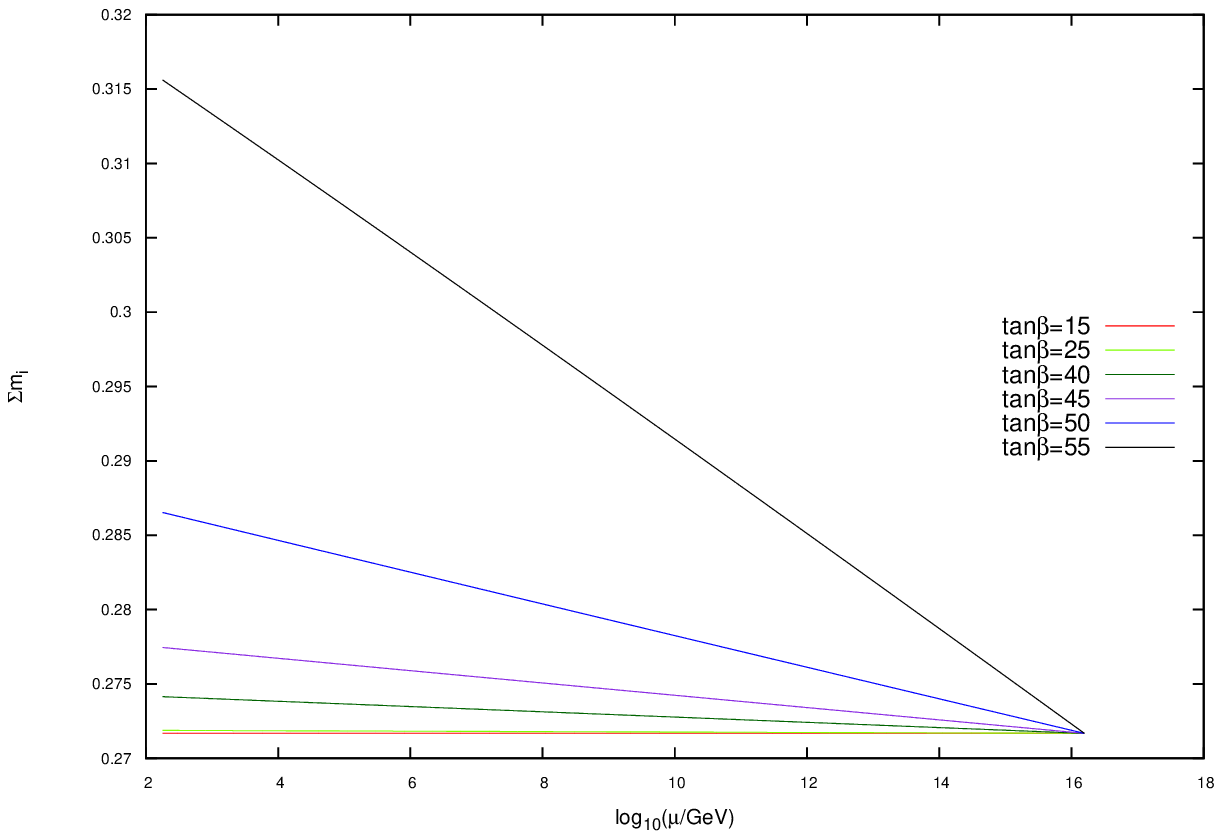}
\caption{Evolution of $\sum_i \lvert m_i \rvert$ for
 tan$\beta$=15, 25, 40, 45, 50, 55 for inverted hierarchy using input values given in Table \ref{values1}}
\label{fig6}
\end{figure}
\end{center}
\begin{center} 
\begin{figure}
\includegraphics[width=1.0\textwidth]{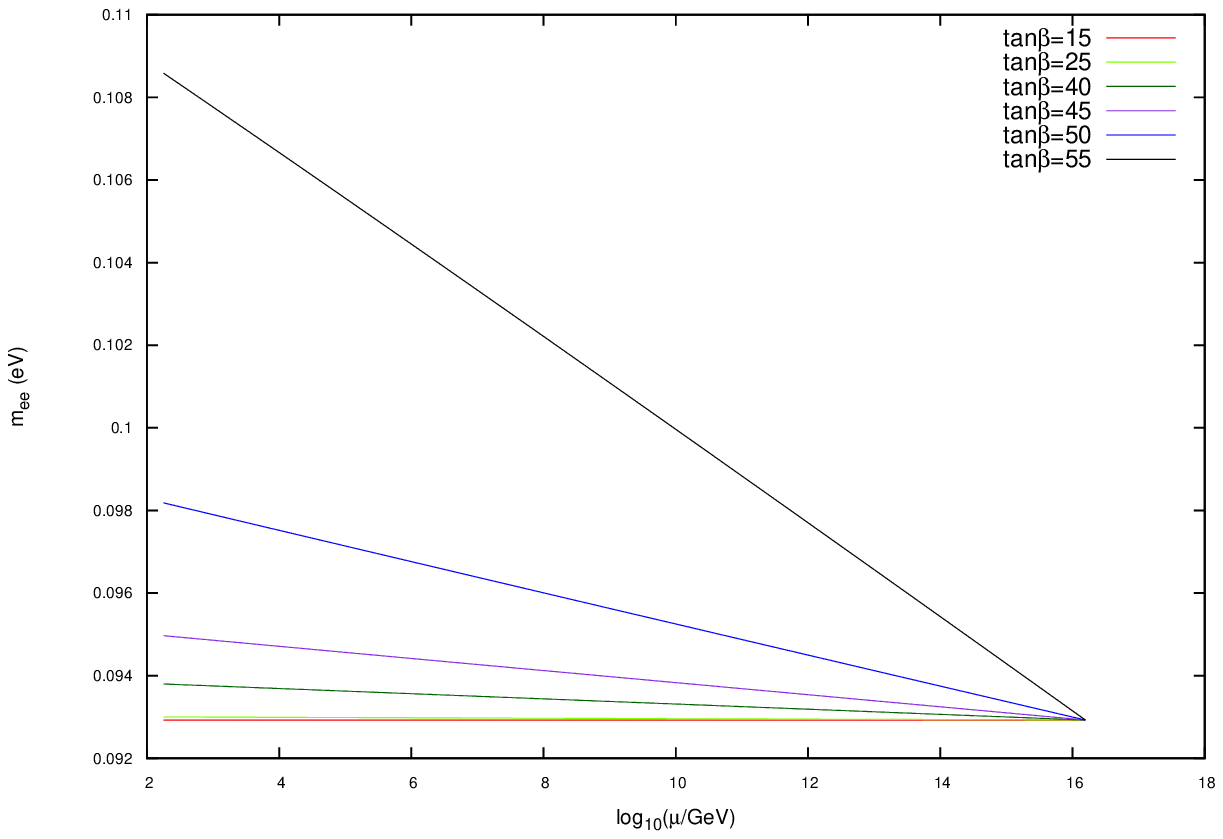}
\caption{Evolution of $m_{ee}$ for
 tan$\beta$=15, 25, 40, 45, 50, 55 for inverted hierarchy using input values given in Table \ref{values1}}
\label{fig7}
\end{figure}
\end{center}
\begin{center} 
\begin{figure}
\includegraphics[width=1.0\textwidth]{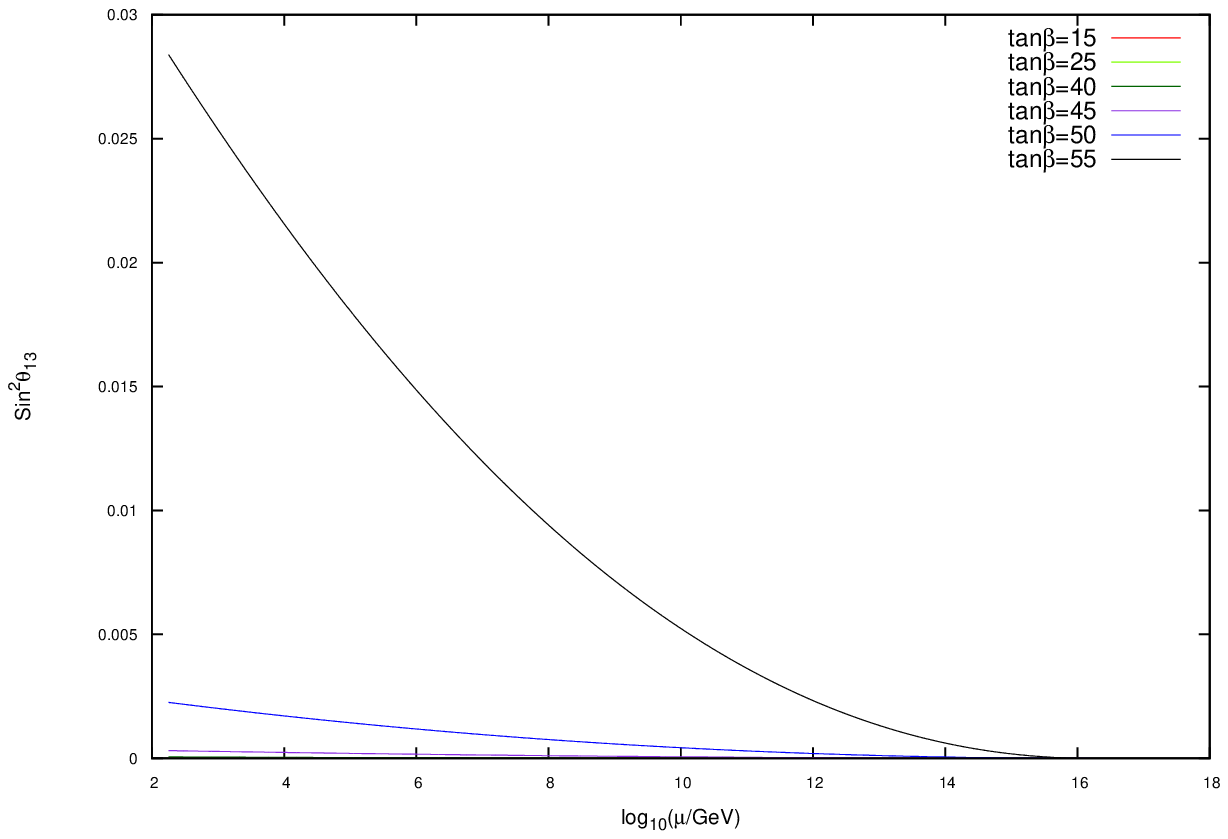}
\caption{Radiative generation of $\text{sin}^2\theta_{13}$ for
 tan$\beta$=15, 25, 40, 45, 50, 55 for normal hierarchy using input values given in Table \ref{values2}}
\label{fig8}
\end{figure}
\end{center}
\begin{center} 
\begin{figure}
\includegraphics[width=1.0\textwidth]{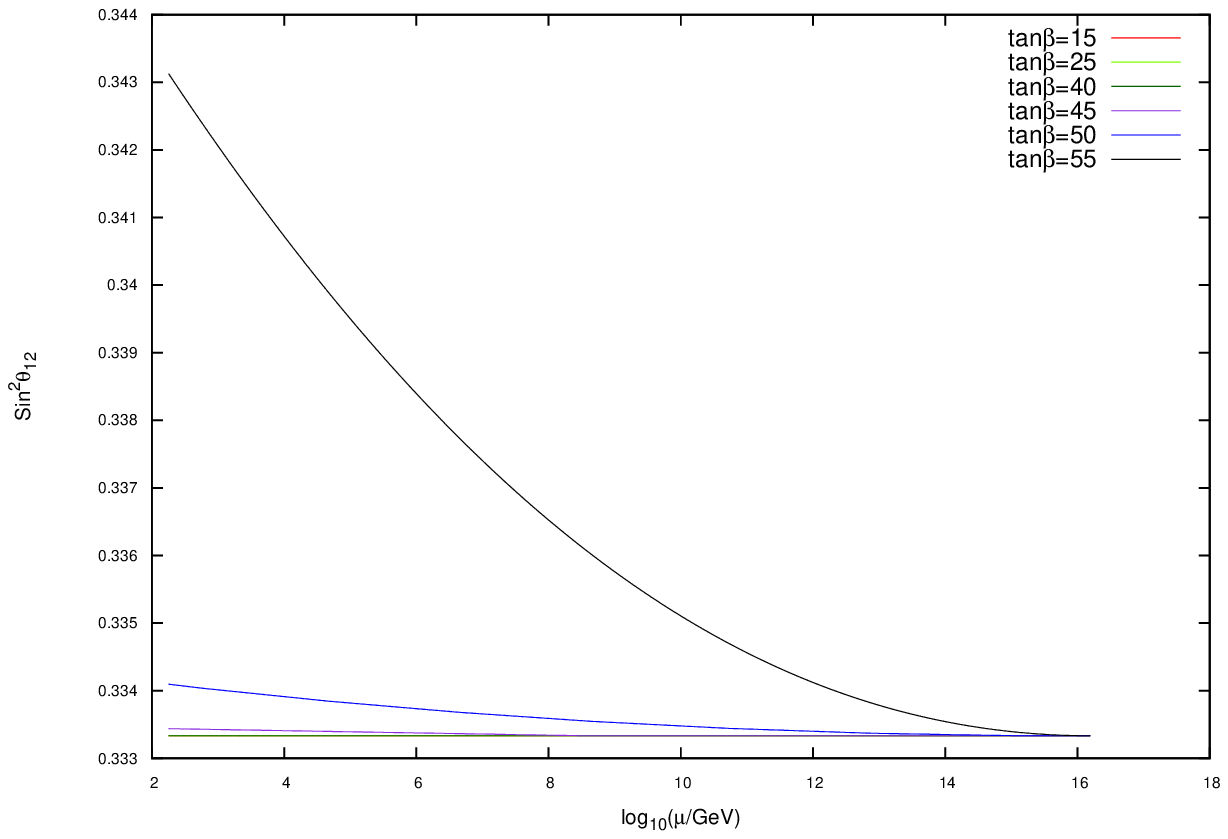}
\caption{Evolution of $\text{sin}^2\theta_{12}$ for
 tan$\beta$=15, 25, 40, 45, 50, 55 for normal hierarchy using input values given in Table \ref{values2}}
\label{fig9}
\end{figure}
\end{center}
\begin{center} 
\begin{figure}
\includegraphics[width=1.0\textwidth]{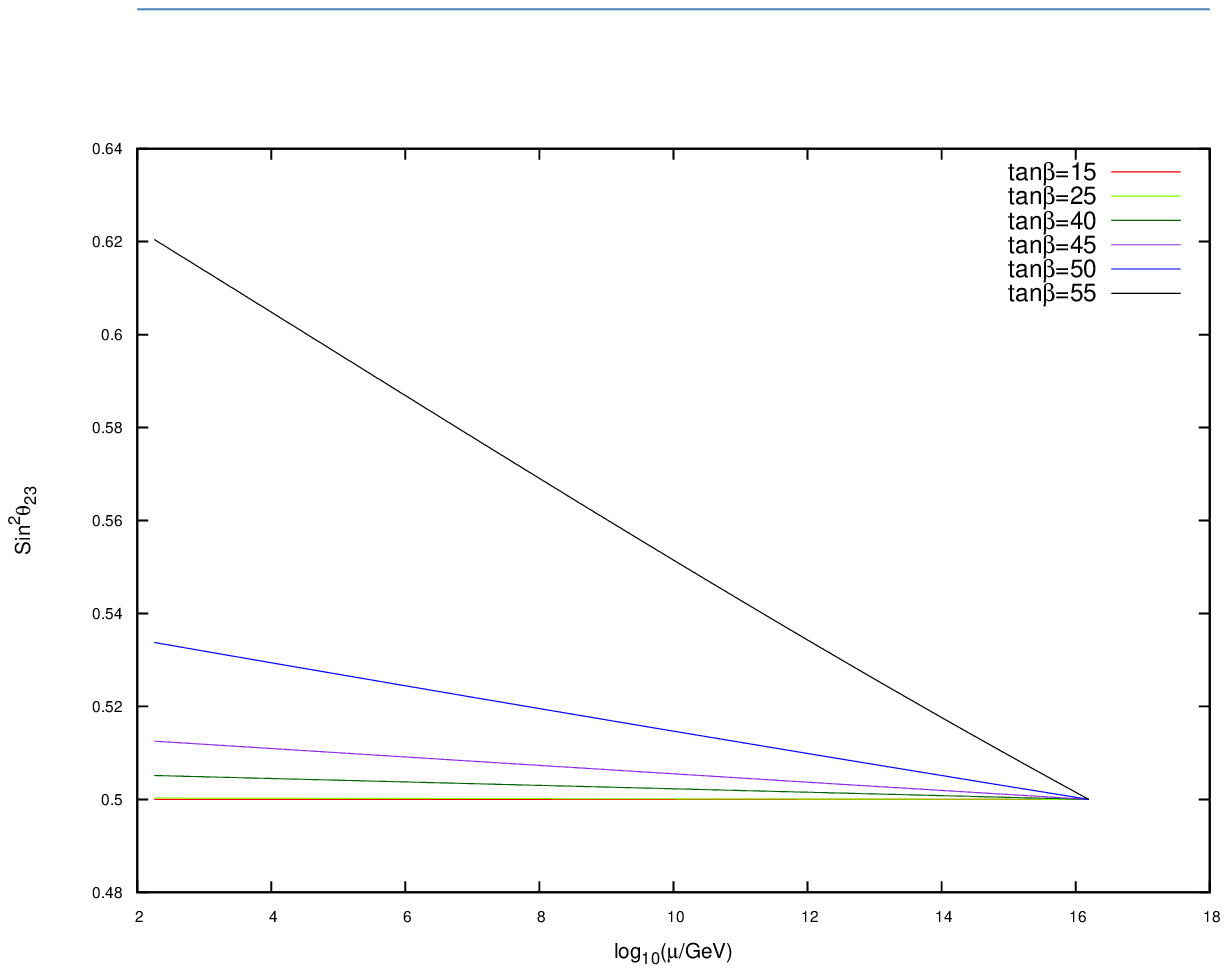}
\caption{Evolution of $\text{sin}^2\theta_{23}$ for
 tan$\beta$=15, 25, 40, 45, 50, 55 using input values given in Table \ref{values2}}
\label{fig10}
\end{figure}
\end{center}
\begin{center} 
\begin{figure}
\includegraphics[width=1.0\textwidth]{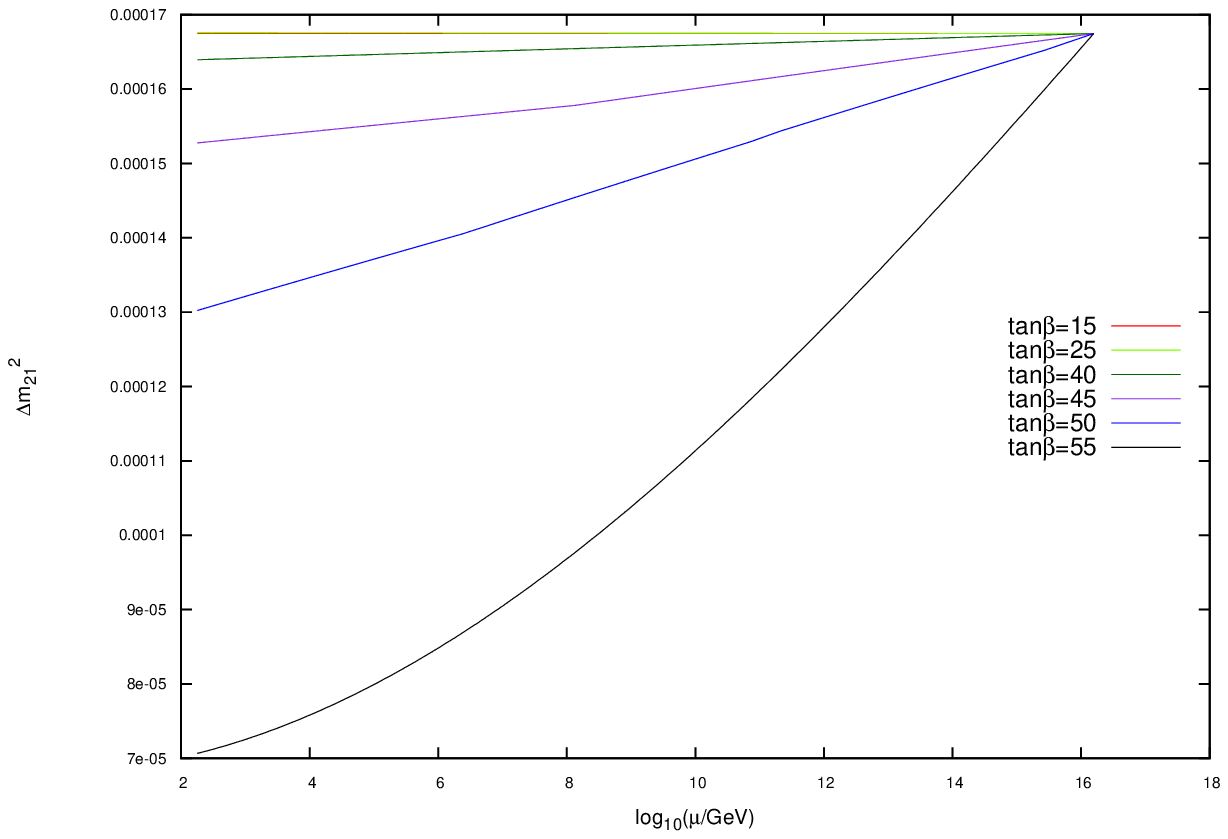}
\caption{Evolution of $\Delta m_{21}^2$ for
 tan$\beta$=15, 25, 40, 45, 50, 55 for normal hierarchy using input values given in Table \ref{values2}}
\label{fig11}
\end{figure}
\end{center}
\begin{center} 
\begin{figure}
\includegraphics[width=1.0\textwidth]{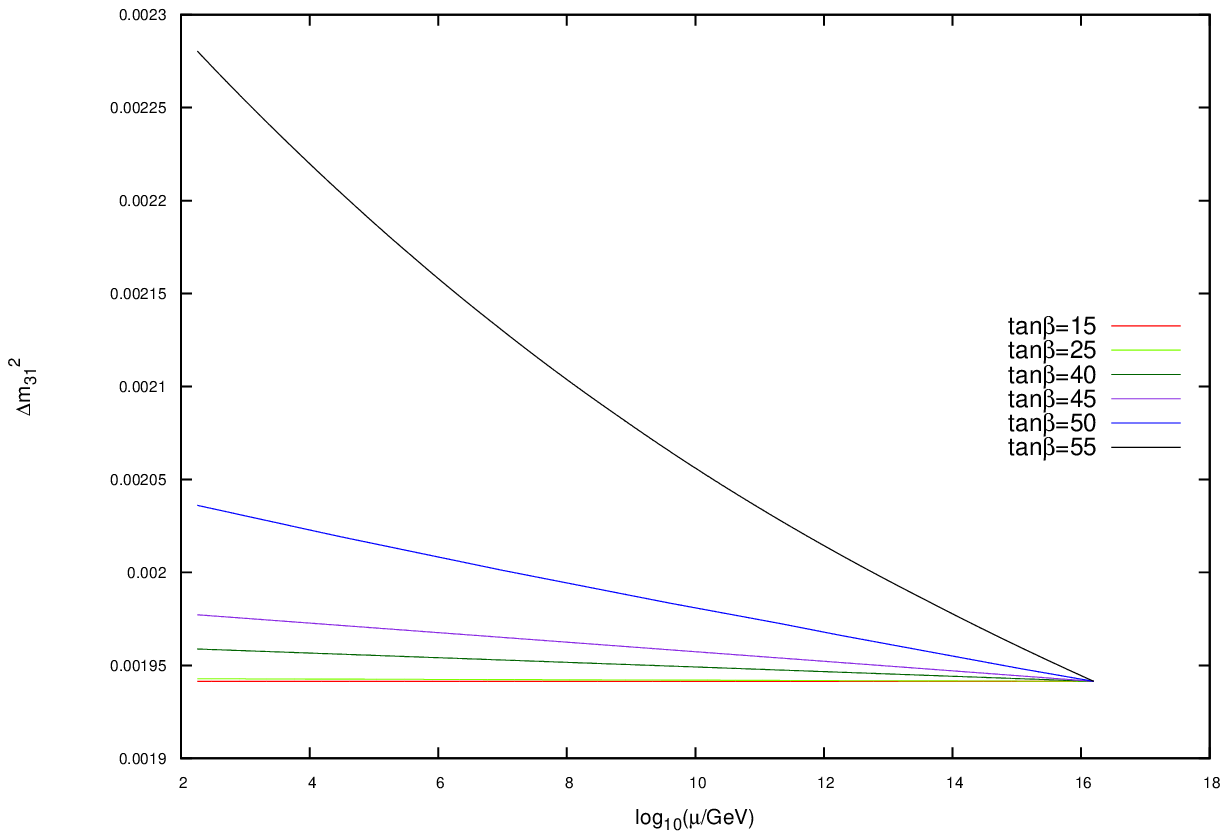}
\caption{Evolution of $\Delta m_{31}^2$ for
 tan$\beta$=15, 25, 40, 45, 50, 55 for normal hierarchy using input values given in Table \ref{values2}}
\label{fig12}
\end{figure}
\end{center}
\begin{center} 
\begin{figure}
\includegraphics[width=1.0\textwidth]{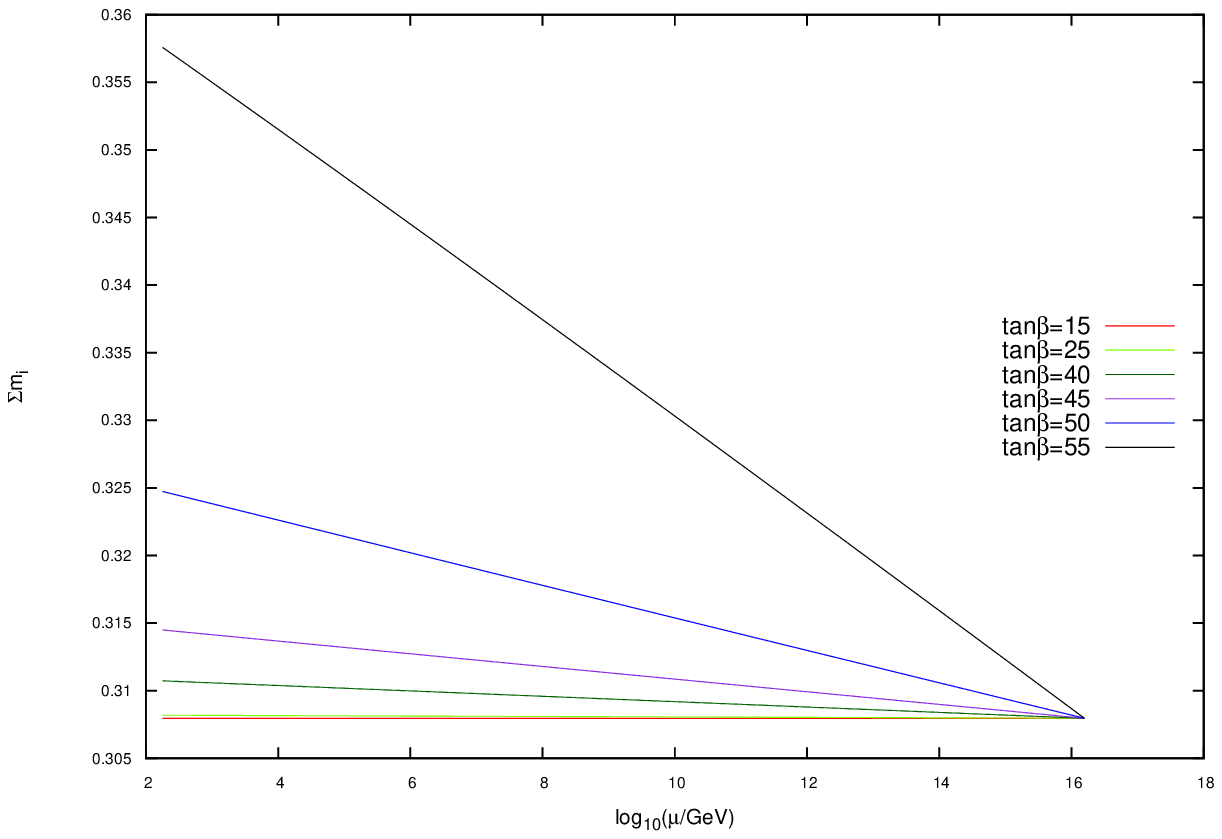}
\caption{Evolution of $\sum_i \lvert m_i \rvert$ for
 tan$\beta$=15, 25, 40, 45, 50, 55 for normal hierarchy using input values given in Table \ref{values2}}
\label{fig13}
\end{figure}
\end{center}
\begin{center} 
\begin{figure}
\includegraphics[width=1.0\textwidth]{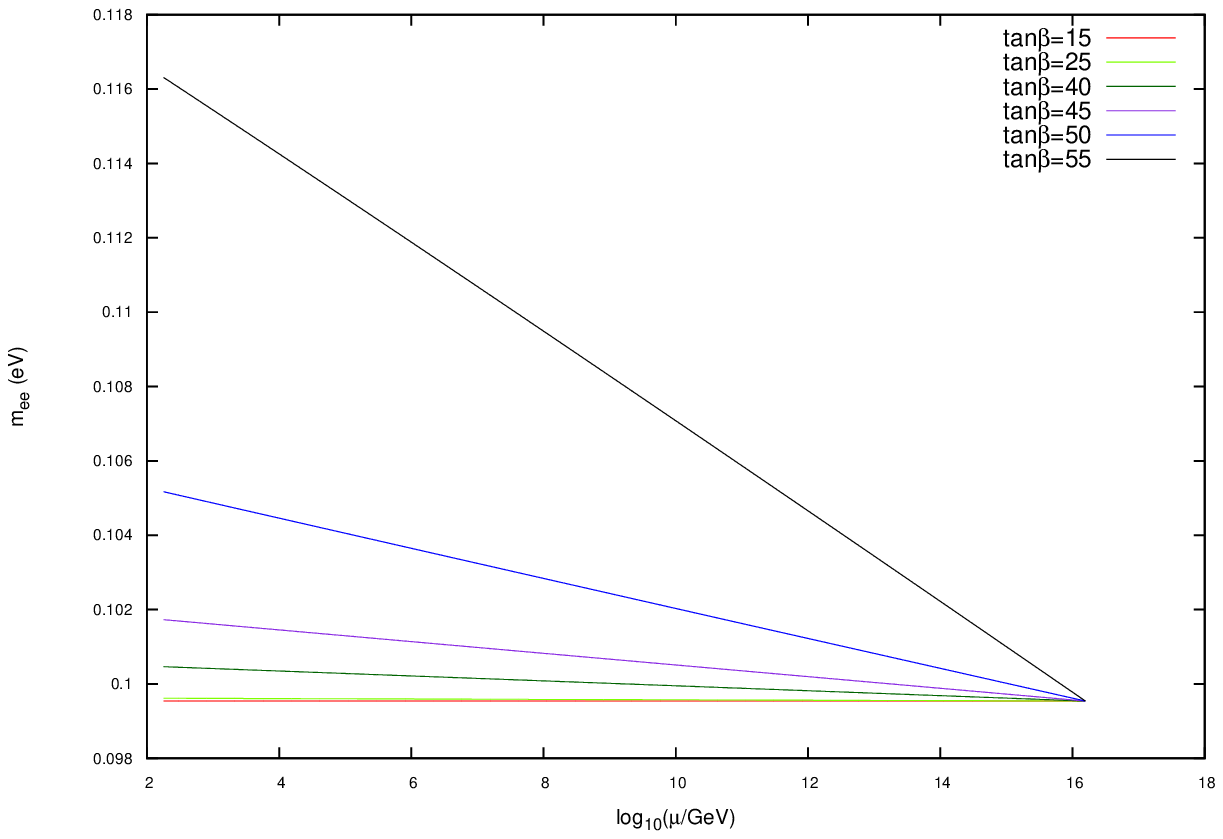}
\caption{Evolution of $m_{ee}$ for
 tan$\beta$=15, 25, 40, 45, 50, 55 for normal hierarchy using input values given in Table \ref{values2}}
\label{fig14}
\end{figure}
\end{center}

\section{Numerical analysis  and results}
\label{result}

\begin{table}[ht]
\begin{center}
\caption{Input and output values with different tan$\beta$ values for Inverted Hierarchy}
\label{values1}
{\small
\begin{tabular}{|p{1.6cm}|p{1.8cm}|p{2.1cm}|p{2.1cm}|p{2.1cm}|p{2.1cm}|p{2.1cm}|p{2.1cm}|}
\hline
    \multicolumn{2}{|c|}{Input Values} & \multicolumn{6}{c|}{Output Values for different tan$\beta$ }              \\
    \hline
   
    --& --                             & tan$\beta$=15 & tan$\beta$=25 &tan$\beta$=40&tan$\beta$=45 & tan$\beta$=50 & tan$\beta$=55                                 \\
    \hline
    $m_1$ (eV) & 0.0924619                             & 0.0924619    & 0.0925375&0.0933433&0.0945343 & 0.0978126 & 0.1086331                                \\
    \hline
    $m_2$ (eV)& -0.0938539                           & -0.0938539 & -0.0939295 &-0.0947101& -0.0958434 &  -0.0989746 & -0.1089959                               \\
    \hline
    $m_3$ (eV) & 0.0853599                             & 0.0853599    & 0.0854102 &0.0860902&0.0870723  & 0.0897417  & 0.0979824                              \\
    \hline
    sin$\theta_{23}$ & 0.707107                            &0.7070999&0.7066970&0.7030523&0.6975724&0.6831660 & 0.6398494                           \\
    \hline
    sin$\theta_{13}$ & 0.00                             &0.0000655&0.0006287&0.0081088&0.0188213&0.0467352 & 0.1265871                              \\
    \hline
    sin$\theta_{12}$& 0.57735                            &0.57735&0.57735&0.57735&0.5774592&0.5779958 & 0.5820936                            \\
    \hline
\end{tabular}
}
\end{center}
\end{table}
\begin{table}[ht]
\centering
\caption{Input and output values with different tan$\beta$ values for Normal Hierarchy}
\vspace{0.5cm}
{\small
\begin{tabular}{|p{1.6cm}|p{1.8cm}|p{2.1cm}|p{2.1cm}|p{2.1cm}|p{2.1cm}|p{2.1cm}|p{2.1cm}|}
\hline
    \multicolumn{2}{|c|}{Input Values} & \multicolumn{6}{c|}{Output Values for different tan$\beta$ }              \\
    \hline
   
    --& --                             & tan$\beta$=15 & tan$\beta$=25 &tan$\beta$=40&tan$\beta$=45 & tan$\beta$=50 & tan$\beta$=55                                 \\
    \hline
    $m_1$ (eV) & 0.0992596                             & 0.0992596    & 0.0993352&0.1001914&0.1014757 & 0.1049422 & 0.1159424                                \\
    \hline
    $m_2$ (eV)& -0.1000997                           & -0.1000996 & -0.1001752 &-0.1010062& -0.1022256 &  -0.1055608 & -0.1162467                               \\
    \hline
    $m_3$ (eV) & 0.1085996                             & 0.1085996    & 0.1086751 &0.1095313&0.1107905  & 0.1142319  & 0.1253917                              \\
    \hline
    sin$\theta_{23}$ & 0.707107                            &0.7070999&0.7073014&0.7107263&0.7159094&0.7305902 & 0.7876961                           \\
    \hline
    sin$\theta_{13}$ & 0.00                             &0.0000582&0.0005604&0.0073647&0.0176199&0.0474922 & 0.1684841                              \\
    \hline
    sin$\theta_{12}$& 0.57735                            &0.57735&0.57735&0.57735&0.5774410&0.5780104 & 0.5857702                            \\
    \hline
\end{tabular}
}
\label{values2}
\end{table}
For the analysis of the RGE's, equations 
(\ref{masseigen}),(\ref{angle1})-(\ref{angle3}) 
for neutrino masses and mixing angles,  
here we follow two consecutive steps (i) bottom-up running \cite{mkp}  in the first place, and then (ii) top-down running \cite{nns} in the next.
 In the first step (i),  the running 
of the RGE's for the third family Yukawa couplings $(h_{t},h_{b}, h_{\tau})$ and three gauge couplings $(g_{1},g_{2},g_{3})$ 
in MSSM , are carried out  from top-quark mass scale ($t_0=\ln m_t$) at low energy end to high energy scale $M_{R}$ \cite{mkp,nns1}.
In the present analysis we consider the high scale value as  the unification scale 
 $M_R=1.6\times 10^{16}$ GeV, with different $\tan\beta$ input values to check the stability of the model at low energy scale. 
For  simplicity of the calculation, the SUSY breaking scale  is taken  at the top-quark 
mass scale $t_0=\ln m_t$ \cite{nns,mkp}.
We adopt the  standard procedure to get the  values of gauge couplings at top-quark mass scale from the experimental CERN-LEP measurements 
at $M_{Z}$, using one-loop RGE's, assuming the existence of a 
one-light Higgs doublet and five quark flavors below $m_t$ scale \cite{mkp,nns1}. Using CERN-LEP data, $M_Z=91.187 GeV$,$ \alpha_s(M_Z)=0.118\pm 0.004$, 
$\alpha^{-1}_1(M_Z)=127.9\pm 0.1$, $sin^2\theta_W(M_Z)=0.2316\pm 0.0003$, and SM relations,\\
\begin{equation}
\frac{1}{\alpha_1(M_Z)}=\frac{3}{5}\frac{(1-sin^2\theta_W(M_Z))}{\alpha(M_Z)}, \frac{1}{\alpha_2(M_Z)}=\frac{sin^2\theta_W(M_Z)}{\alpha(M_Z)}  \ \ 
,  g_i^2=4\pi\alpha_i,
\end{equation}
 
we calculate the gauge couplings at $M_Z$ scale, $\alpha_1(M_Z)=0.0169586$, $\alpha_2(M_Z)=0.0337591$, $\alpha_3(M_Z)=0.118$. As already mentioned, 
we consider the existence of one light Higgs doublet $(n_H=1)$ and five quark flavors $(n_F=5)$ in the scale $M_Z-m_t$. Using one-loop RGE's of gauge 
couplings, we get $g_1(m_t)=0.463751$, $g_2(m_t)=0.6513289$ and $g_3(m_t)=1.1891996$. Similarly, the Yukawa couplings are also evaluated at top-quark mass 
scale for input values of  $m_t(m_t)=174$ GeV, $m_b(m_b)=4.25$ GeV, $m_{\tau}(m_{\tau})=1.785$ GeV and the QED-QCD rescaling factors
$\eta_{b}=1.55$, $\eta_{\tau}=1.015$ in the standard fashion \cite{nns1},\\

$$h_t(m_t)=\frac{m_t(m_t)\sqrt{1+tan^2\beta}}{174  tan\beta},$$
$$h_b(m_t)=\frac{m_b(m_t)\sqrt{1+tan^2\beta}}{174},$$
\begin{equation}
h_{\tau}(m_t)=\frac{m_{\tau}(m_t)\sqrt{1+tan^2\beta}}{174}.
\end{equation} 
where $m_b(m_t)=\frac{m_b(m_b)}{\eta_b}$, $m_{\tau}(m_t)=\frac{m_\tau(m_{\tau})}{\eta_{\tau}}$. The one-loop RGE's
for top quark, bottom quark and $\tau$-lepton Yukawa couplings in the MSSM in the range of mass scales $m_t\leq\mu\leq M_R$ are given by
\begin{equation}
\frac{d}{dt}h_t=\frac{h_t}{16\pi^2}(6h_t^2+h_b^2-\sum^3_{i=1}c_ig^2_i),
\label{h1}
\end{equation}

\begin{equation}
\frac{d}{dt}h_b=\frac{h_b}{16\pi^2}(6h_b^2+h_{\tau}^2+-\sum^3_{i=1}c_i^{'}g_i^2),
\label{h2}
\end{equation}

\begin{equation}
\frac{d}{dt}h_{\tau}=\frac{h_{\tau}}{16\pi^2}(4h_{\tau}^2+3h_b^2-\sum^3_{i=1}c^{''}_ig^2_i),
\label{h3}
\end{equation}
where
\begin{equation}
c_i=\left(\begin{array}{c}\frac{13}{15}\\3\\ \frac{16}{3}\end{array}\right),  
c_i^{'}=\left(\begin{array}{c}\frac{7}{15}\\3\\ \frac{16}{3}\end{array}\right), 
c_i^{''}=\left(\begin{array}{c}\frac{9}{5}\\3\\ 0\end{array}\right).
\end{equation}
The two-loop RGE's for the gauge couplings are similarly expressed in the range of mass scales $m_t\leq\mu\leq M_R$ as
\begin{equation}
\frac{d}{dt}g_i=\frac{g_i}{16\pi^2}[b_ig_i^2+\frac{1}{16\pi^2}(\sum_{j=1}^3b_{ij}g_i^2g_j^2)-\sum_{j=t, b, \tau}^3a_{ij}g_i^2h_j^2)],
\label{g}
\end{equation}
where
\begin{equation}
b_i=\left(\begin{array}{c}6.6\\1\\ -3\end{array}\right), 
b_{ij}=\left(\begin{array}{ccc}7.9 & 5.4 & 17\\1.8 & 25 & 24\\ 2.2 & 9 & 14\end{array}\right), 
a_{ij}=\left(\begin{array}{ccc}5.2 & 2.8 & 3.6\\6 & 6 & 2\\ 4 & 4 & 0\end{array}\right).
\end{equation}
Values of $h_t$, $h_b$, $h_{\tau}$, $g_1$, $g_2$, $g_3$  evaluated for tan$\beta = 55$ at high scale $M_R=1.6\times 10^{16}$ from equation (\ref{h1})-(\ref{h3}) and (\ref{g}) are \\
\begin{center}
$h_t(M_R)=0.142685458,  h_b(M_R)=0.378832042$\\
$h_{\tau}(M_R)=0.380135357,  g_1(M_R)=0.381783873$\\
$g_2(M_R)=0.377376229,  g_3(M_R)=0.374307543$
\end{center}
In the second step (ii), the running of three neutrino masses $(m_1, m_2, m_3)$ and mixing angles $(s_{12}, s_{23}, s_{13})$ are carried out together with
the running of Yukawa and gauge couplings, from high scale $t_R(=ln M_R)$ to low scale $t_o$. In this case, we use the input values of Yukawa and gauge 
couplings evaluated earlier at scale $t_R$ from the first stage running of RGE's in case (i). In principle, one can evaluate neutrino masses and mixing 
angles at every point of the energy scale.  It can be noted that in the present problem, the running of other SUSY parameters such 
 as $M_0$, $M_{1/2}$, $\mu$, are not required and hence, it is not necessary to supply their input values. 
 
We are now interested in studying radiative generation  $\theta_{13}$ for the  case when $m_{1,2,3} \neq 0$ and $s_{13}=0$ at high energy scale. 
 Such  studies can give the possible origin of the reactor angle in a broken $A_4$ model. 
  During the running  of mass eigenvalues and mixing angles from high to low scale, 
the non-zero input  value of mass eigenvalues $m_{1,2,3}$ will induce radiatively  a  non-zero values of $s_{13}$. Similar approach was followed 
in \cite{anjan} considering $m_3=0$. The authors in \cite{anjan} used inverted hierarchy neutrino mass pattern $(m, -m, 0)$ at high scale. Such a specific structure of mass eigenvalues however, require fine tuning conditions in the flavor symmetry model at high energy. Instead of assuming a specific relation between mass eigenvalues at high energy scale, here we attempt to find out the most general mass eigenvalues at high energy which can give rise to the correct neutrino data at low energy scale. The only assumption in our
work is the opposite CP phases i.e. $(m_1, -m_2, m_3)$. In another work \cite{aun}, 
authors have shown the radiative generation of $\bigtriangleup m^2_{21}$ considering the non-zero $\theta_{13}$ at high scale and  tan$\beta$ 
values lower than 50. They have also shown that $\Delta m^2_{21}$  can run from zero at high energy to the observed value at the low energy scale, only 
if $\theta_{13}$ is relatively large and the Dirac CP-violating phase is close to $\pi$. The running effects can be observed 
only when $\theta_{13}$ is non-zero at high-energy scale as per their analysis. In the present work, $\theta_{13}$ is assumed to be zero at high scale consistent with a TBM type mixing within $A_4$ symmetric model. We also examine the 
running behavior of neutrino parameters in a neutrino mass model obeying special kind of $\mu$-$\tau$ symmetry at high scale, which was not studied in the earlier work 
mentioned above. 

For a complete numerical analysis, first we parameterize the neutrino mass matrix to have a TBM type structure with eigenvalues in the form ($m_1, -m_2, m_3$). Since the mixing angles at high energy scale are fixed (TBM type), we only need to provide three input values namely, $m_1, m_2, m_3$. Using these values at the high energy scale, neutrino 
parameters are computed at low energy scale by simultaneously solving the RGE's discussed above. We first allow moderate as well as large hierarchies between the lightest and the heaviest mass eigenvalues (with the lighter being at least two orders of magnitudes smaller) of both normal and inverted type and find that the output values of $\theta_{13}$ do not lie in the experimentally allowed range for all values of $\tan{\beta} = 15, 25, 40, 45, 50, 55$ used in our analysis. 

We then consider very mild hierarchical pattern of mass eigenvalues keeping them in the same order of magnitude range. We vary the neutrino mass eigenvalues at high energy scale in the range $0.01-0.12$ eV and generate the neutrino parameters at low energy. We restrict the neutrino parameters $\theta_{13}, \theta_{12}, \theta_{23}$ and $\Delta m_{21}^2$ at low energy to be within the allowed $3\sigma$ range and show the variation of $\Delta m_{23}^2 (\text{IH}), \Delta m_{31}^2 (\text{NH})$ at low energy with respect to the input mass eigenvalues at high energy. We show the results in figure \ref{fig15}, \ref{fig16}, \ref{fig17}, \ref{fig18}, \ref{fig19} and \ref{fig20} for a specific value of tan$\beta=55$. It can be seen from these figures that the correct value of neutrino parameters at low energy can be obtained only for large values of mass eigenvalues at high energy scale $|m_{1,2,3}| = 0.08-0.12$ eV. We then choose two specific sets of mass eigenvalues at high energy scale corresponding to inverted hierarchy and normal hierarchy respectively and show the evolution of several neutrino observables including oscillation parameters, effective neutrino mass $m_{ee} = \lvert \sum_i U^2_{ei} m_i \rvert$, sum of absolute neutrino masses $\sum_i \lvert m_i \rvert$ in figure \ref{fig1}, \ref{fig2}, \ref{fig3}, \ref{fig4}, \ref{fig5}, \ref{fig6}, \ref{fig7}, \ref{fig8}, \ref{fig9}, \ref{fig10}, \ref{fig11}, \ref{fig12}, \ref{fig13}, \ref{fig14}. It can be seen from figure \ref{fig1} and \ref{fig8} that the correct value of $\theta_{13}$ can be obtained at low energy only for very high values of $\tan{\beta} = 55$. The other neutrino parameters also show a preference for higher $\tan{\beta}$ values. The output values of neutrino parameters at low energy are given in table \ref{values1} and \ref{values2} for both sets of input parameters. The large deviation of $\theta_{13}$ at low energy from its value at high energy ($\theta_{13}=0$ for TBM at high energy) whereas smaller deviation of other two mixing angles can be understood from the RGE equations for mixing angles (\ref{angle1}), (\ref{angle2}), (\ref{angle3}). Using the input values given in table \ref{values1} and \ref{values2}, the slope of sin$\theta_{13}$ can be calculated to be $\frac{h^2_{\tau}}{16\pi^2}(-5.88)$ and $\frac{h^2_{\tau}}{16\pi^2}(5.23)$ for inverted and normal hierarchies respectively. On the other hand, the slope of sin$\theta_{23}$ at high energy scale is found to be $\frac{h^2_{\tau}}{16\pi^2}(2.95)$ and $\frac{h^2_{\tau}}{16\pi^2}(-2.63)$ for inverted and normal hierarchies respectively. Thus, the lower value of slope for sin$\theta_{23}$ results in smaller deviation from TBM values compared to that of sin$\theta_{13}$. We also note from figure \ref{fig6}, \ref{fig13} that the sum of the absolute neutrino masses at low energy is $0.315$ eV and $0.3555$ eV for inverted and normal hierarchy respectively. This lies outside the limit set by the Planck experiment $\Sigma |m_i| < 0.23$ eV \cite{planck}. However, there still remains a little room for the sum of absolute mass to lie beyond this limit depending on the cosmological model, as suggested by several recent studies \cite{nucosmo}. Ongoing as well as future cosmology experiments should be able to rule out or confirm such a scenario.

It is interesting to note that, our analysis shows a preference for very mild hierarchy of either inverted or normal type at high energy scale which also produces a very mild hierarchy at low energy. This can have interesting consequences in the ongoing neutrino oscillation as well as neutrino-less double beta decay experiments. Also, the large $\tan{\beta}$ region of MSSM (which gives better results in our model) will undergo serious scrutiny at the collider experiments making our model falsifiable both from neutrino as well as collider experiments. We note that the present analysis will be more accurate if the two loop contributions \cite{twlp} RGE's are taken into account.



\section{Conclusion}\label{conclude}
We have studied the effect of RGE's on neutrino masses and mixing in MSSM with $\mu-\tau$ symmetric neutrino mass model giving TBM type mixing at high energy scale. We incorporate an additional flavor symmetry $A_4$ at high scale to achieve the desired structure of the neutrino mass matrix. The RGE equations for different neutrino parameters are numerically solved simultaneously for different values of $\tan{\beta}$ ranging from 15 to 55. We take the three neutrino mass eigenvalues at high energy scale as free parameters and determine the parameter space that can give rise to correct values of neutrino oscillation parameters at low energy. We make the following observations

 \begin{itemize}
 \item Moderate or large hierarchy (both normal and inverted) of neutrino masses at high energy scale does not give rise to correct output at low energy scale.
 \item Very mild hierarchy (with all neutrino mass eigenvalues having same order of magnitude values and $|m_{1,2,3}| = 0.08-0.12$ eV) give correct results at low energy provided the $\tan{\beta}$ values are kept high, close to 55. Such a preference towards large mass eigenvalues with all eigenvalues having same order of magnitude values can have tantalizing signatures at oscillation as well as neutrino-less double beta decay experiments.
  \item No significant changes in running of $\sin^2\theta_{23}$, $\sin^2\theta_{12}$ with $\tan{\beta}$ are observed.  
   
  \item Sum of absolute neutrino masses at low energy lie above the Planck upper bound $\Sigma |m_i| < 0.23$ eV \cite{planck} hinting towards non-standard cosmology to accommodate a larger $\Sigma |m_i|$ or more relativistic degrees of freedom \cite{nucosmo}.
  
  \item The preference for high $\tan{\beta}$ regions of MSSM could go through serious tests at collider experiments pushing the model towards verification or falsification.
  \end{itemize}
  
Although we have arrived at some allowed parameter space in our model giving rise to correct phenomenology at low energy with the additional possibility that many or all of these parameter space might get ruled out in near future, we also note that it would have been more interesting if the running of the Dirac and Majorana CP violating phases \cite{lx} were taken into account. We also have not included the seesaw threshold effects and considered all the right handed neutrinos to decouple at the same high energy scale. Such threshold effects could be important for large values of $\tan{\beta}$ as discussed in \cite{threshold}. We leave such a detailed study for future investigations.

\section{Acknowledgement}
The work of MKD is partially supported by the grant no. 42-790/2013(SR) from UGC, Govt. of India.

\end{document}